\documentclass[
  journal=pasa,
  manuscript=research-paper, %% or "review"
  year=2020,
  volume=37,
]{cup-journal}

\usepackage{microtype,siunitx,booktabs,amsmath}
\usepackage{ulem}
\title{Non-gaussianity of optical emission lines in SDSS star-forming galaxies and its implications on galactic outflows}
\author{B. P. Brian Yu}
\affiliation{Mullard Space Science Laboratory, University College London, Holmbury St Mary, Surrey, RH5 6NT, UK}
\email[B. P. Brian Yu]{brian.yu.16@ucl.ac.uk}

\author{James Angthopo}
\affiliation{Mullard Space Science Laboratory, University College London, Holmbury St Mary, Surrey, RH5 6NT, UK}
\alsoaffiliation{INAF, Osservatorio Astronomico di Brera, Via Brera 28, 20121, Milano, Italy}

\author{Ignacio Ferreras}
\affiliation{Instituto de Astrof\'isica de Canarias, C/V\'ia L\'actea, s/n, E38205 La Laguna, Tenerife, Spain}
\alsoaffiliation{Department of Physics and Astronomy, University College London, Gower Street, London, WC1E 6BT, UK}
\alsoaffiliation{Departamento de Astrof{\'i}sica, Universidad de La Laguna (ULL), E-38206 La Laguna, Tenerife, Spain}

\author{Kinwah Wu}
\affiliation{Mullard Space Science Laboratory, University College London, Holmbury St Mary, Surrey, RH5 6NT, UK}
\alsoaffiliation{Research Center for Astronomy, Astrophysics and Astrophotonics, Macquarie University, Sydney, NSW 2019, Australia}

\doi{10.1017/pasa.2020.32}

\received {dd Mmm YYYY}
\revised  {dd Mmm YYYY}
\accepted {dd Mmm YYYY}
\published{22 September 2020}

\begin{document}

\begin{abstract}
The shape of emission lines in the optical spectra of star-forming galaxies reveals the kinematics of the diffuse gaseous component. We analyse the shape of prominent emission lines in a sample of $\sim$53,000 star-forming galaxies from the Sloan Digital Sky Survey, focusing on departures from gaussianity. Departures from a single gaussian profile allow us to probe the motion of gas and to assess the role of outflows. The sample is divided into groups according to their stellar velocity dispersion and star formation rate. The spectra within each group are stacked to improve the signal-to-noise ratio of the emission lines, to remove individual signatures, and to enhance the effect of star formation rate on the shapes of the emission lines. The moments of the emission lines, including kurtosis and skewness, are determined. We find that most of the emission lines in strong star-forming systems unequivocally feature negative kurtosis. This signature is present in H$\beta$, H$\alpha$, [\textsc{N\,ii}] and [\textsc{S\,ii}] in massive galaxies with high star formation rates. We attribute it as evidence of radial outflows of ionised gas driven by the star formation of the galaxies. Also, most of the emission lines in low-mass systems with high star formation rates feature negative skewness, and we interpret it as evidence of dust obscuration in the galactic disk. These signatures are however absent in the [\textsc{O\,iii}] line, which is believed to trace a different gas component. The observed trend is significantly stronger in face-on galaxies, indicating that star formation drives the outflows along the galactic rotation axis, presumably the path of least resistance. The data suggest that outflows driven by star formation exert accumulated impacts on the interstellar medium, and the outflow signature is more evident in older galaxies as they have experienced a longer total duration of star formation.
\end{abstract}

\section{INTRODUCTION}

Star-formation is an important process driving galactic evolution, produced by the collapse and cooling of gas towards the gravitational potential wells of dark matter halos. This collapse is either a result of gas infall onto these halos, or triggered by encountering events, such as galaxy mergers and tidal interactions with nearby galaxies that can induce a burst of star formation. However, galaxies are not a homogeneous class of stellar systems, resulting in a variety of environments, where stars are formed at different rates, even for the same amount of gas inflow. Survey observations, such as the Sloan Digital Sky Survey \citep[hereafter SDSS,][]{York2000, Gunn2006}, have shown a tight correlation between the star formation rate (SFR) and the galactic stellar mass \citep{Lilly2013,Speagle2014}, which manifests as the Main Sequence of star-forming galaxies.

The interstellar gas exists as a multi-phase medium, where the different gaseous components can be constrained over a wide spectral range, from X-ray to infrared. This paper focuses on optical spectra, mostly contributed by the stellar population -- as a continuum along with a complex network of absorption lines -- and gas in various stages of ionisation. The luminosity of key emission lines, such as H$\alpha$ or [\textsc{O\,ii}], are used as proxies of the star formation rate \citep{Kennicutt1998}, and ratios between the luminosity of recombination and collisional lines allow us to constrain the physical properties of the gas, including density, temperature, as well as discriminating the ionisation state of the emitting gas as being either star-forming, AGN, or shock-driven \citep{Baldwin1981, Kewley2001, Peimbert2017, Kewley2019}. The complexity of the gas distribution in the interstellar medium, with material having different physical properties (pressure, temperature, density, ionisation state, etc.) and moving at different velocities, implies that the gas kinematics will leave different signatures on different types of lines \citep[see, e.g.,][]{Osterbrock2006,Tanner2017}.

An episode of star formation could typically last for hundreds of Myr \citep{DiMatteo2008}. During the starburst episode, the explosions of the massive stars at the end of their life cycles release a large amount of energy, which disrupts the gas inflow and also dispels the gas from the central regions of the galaxies, and could launch a galactic-scale outflow \citep{Veilleux2005}. The outflow depends on the activity, size, and shape of the star-forming region, and, in turn, the galactic outflow would influence and even quench the subsequent star-forming process \citep[see][and references therein]{Veilleux2005, Kewley2019}. Star-formation feedback is assumed to drive the transition of galaxies towards quiescence at the faint end, with star-forming systems being the dominant fraction in the Green Valley \citep[see, e.g.,][]{Salim2014,Angthopo2019}, and it is also invoked as the cause of the decreasing stellar to dark matter fraction towards lower mass galaxies in abundance matching studies \citep{Behroozi2010}. Moreover, galaxy mergers can also develop a galactic wind by shock heating the gas \citep{Cox2004, Martin2006}. Lower-mass galaxies inhabit shallower gravitational potentials, thus allowing the outflowing material, which is enriched with metals, to escape the galaxy after interacting and mixing with the star-forming gas \citep{Dekel1986, Martin1999, Ferrara2000, Tremonti2004, Dave2011}. Therefore, the characterisation of gas kinematics and its connection to outflows is of paramount importance.

For spectra at sufficiently high resolution, the emission-line profiles can be resolved to provide additional information about the gas kinematics. For instance, the Na\,I D absorption lines trace neutral gas that is seen in absorption against the background starlight. Under the presence of outflows, the gas is accelerated towards the observer and the Na\,I D lines are blueshifted with respect to the systematic velocity of the host galaxy \citep{Heckman2000, Rupke2002, Martin2005}, which can be measured to estimate the outflow velocity of neutral gas. Double-peaked emission line profiles also indicate bipolar distribution of outflow perpendicular to the disk, where the gap between the double peaks represent the velocity difference in the red and blue wings \citep[e.g.][where the split is measured to be $\sim$1500~km s$^{-1}$ in NGC 3079]{Veilleux1994}. The advent of large integral field unit (IFU) surveys such as MaNGA \citep{Bundy2015} and SAMI \citep{Croom2012} also enabled a more detailed analysis of the radial trends of the outflow signature. \cite{Roberts2020} explore the Na\,I D lines in a sample of Main Sequence star-forming galaxies to find a substantial fraction ($\sim20\%$) showing signatures of outflows, especially in galaxies with a high surface density of the star formation rate, with a strong declining radial trend, so that outflows are stronger in the central, denser regions of star-formation activity. The kinematics of extraplanar gas can be measured in IFU data to assess the presence of outflowing gas, as presented in \citet{Ho2016}, who find wind signatures in galaxies with a high SFR density, results that are consistent with theoretical models from the EAGLE hydrodynamical simulations \citep{Tescari2018}. These resolved studies are also capable of confirming the signature of shock heated gas, as expected in an outflowing scenario \citep{Ho2014}. While single fibre measurements -- as in this paper -- lack the spatial discrimination of IFU data, the asymmetries found in resolved studies motivate the search for departures of emission line profiles from the standard Gaussian function in the standard SDSS spectra. Note that environment may also affect the asymmetry of the emission line profiles \citep{Bloom2018}. However, an analysis based on a large volume of data plotted against SFR would indeed confirm the connection between line shape variations and the presence of outflows.

Outflows from individual galaxies can only be probed when the star-formation activity is extremely efficient \citep[e.g.][who focused on a sample of ULIRGs]{Chen2010}. It is therefore useful to stack spectra of galaxies with similar properties to improve the signal to noise ratio (S/N) so that faint signatures of outflows can be detected in galaxies with mild star-formation activity. This requires data from large galaxy surveys such as the SDSS, which not only allows the S/N to be enhanced significantly with the sheer amount of galaxies detected but also covers a large range of galactic properties. For example, \cite{Chen2010} analysed a sample of star-forming galaxies from SDSS to investigate how the properties of Na\,I D absorption lines vary against galaxies of different inclination angle, specific SFR (sSFR), dust extinction ($A_V$), or stellar mass. They found that the Na\,I D absorption lines are made up of a disk component and an outflowing gas component, which can be detected in edge-on and face-on galaxies, respectively. The gas component has an opening angle of $\sim60^\circ$, perpendicular to the disk, and the Na\,I D equivalent width of gas depends mainly on the sSFR and secondarily on $A_V$, which may be related to the amount of absorbing gas and its survival rate, respectively. \cite{Cicone2016} analysed the emission lines H$\alpha$, [\textsc{N\,ii}], and [\textsc{O\,iii}] in a sample of SDSS star-forming galaxies to investigate how the line properties vary with respect to stellar mass and SFR simultaneously. They characterised the outflow velocity ($v_{\rm out}$) by using the difference between the high-velocity tail of gas and stellar kinematics, and found that $v_{\rm out}$ scales with SFR when SFR $>1\rm\ M_\odot\ yr^{-1}$, whereas the scaling is nearly flat when SFR $<1\rm\ M_\odot\ yr^{-1}$. \cite{Concas2017} analysed a sample of galaxies from SDSS to investigate how the properties of the [\textsc{O\,iii}] emission line vary against different stellar mass, SFR, and BPT classification. They found that the [\textsc{O\,iii}] line profiles of star-forming galaxies are symmetric and narrow, and concluded that there is no significant evidence for starburst-driven outflows in the global population. Instead, an additional blueshifted and broad component is found in the [\textsc{O\,iii}] line of active galaxies, which shows that AGN is responsible for driving strong bulk motion in the warm ionised gas. \cite{Chen2016} analysed a sample of disc star-forming galaxies from SDSS and found that a significant fraction of this sample contains H$\alpha$ emission line with negative kurtosis. Such fraction depends mainly on the the stellar mass and secondarily on the sSFR. Since the fraction is larger in edge-on systems than in face-on systems, they concluded that their findings can be interpreted as a result of rotating galaxy disk with a ring-like H$\alpha$ emission region. We note that the imprint on emission lines from AGN-driven outflows and star-formation-driven outflows will not be the same, as the former corresponds to an injection of energy and momentum within a comparatively smaller region, whereas the energy input from star-formation extends over much larger volumes.

The aim of this paper is to investigate starburst-driven galactic outflows by analysing the shape of the emission lines H$\beta$, [\textsc{O\,iii}], [\textsc{N\,ii}], H$\alpha$, and [\textsc{S\,ii}] in star-forming galaxies. In particular, we quantify the outflow velocity by measuring their kurtosis. Our sample of galaxies is drawn from the SDSS, and the selection and sampling methods are summarised in \S\ref{sec:data}. We describe our stacking procedure, stellar continuum fitting method, as well as the emission line model which quantifies the presence of outflows in \S\ref{sec:process}. The emission line properties and their variations against different galactic properties are presented in \S\ref{sec:result}, and we discuss the outflow properties and how our results compare with those from other works in \S\ref{sec:discuss}. Our findings are summarised in \S\ref{sec:conclude}.

\section{OBSERVATIONAL DATA}
\label{sec:data}

\subsection{Galactic spectra}
\label{sec:spec}

From the SDSS \citep{York2000} Data Release 14 \citep{Abolfathi2018}, we retrieved galaxy spectra directly from the Main Galaxy Sample \citep{Strauss2002}, with Petrosian r-band magnitude of $14.5<r_{\rm AB}<17.7$. These spectra were taken through the 3$^{\prime\prime}$ diameter fibres, and the wavelength ranges from 3800 \AA{} to 9200 \AA{} with resolution increasing from 1500 to 2500, respectively \citep{Smee2013}.

\begin{figure}
\includegraphics[width=\columnwidth]{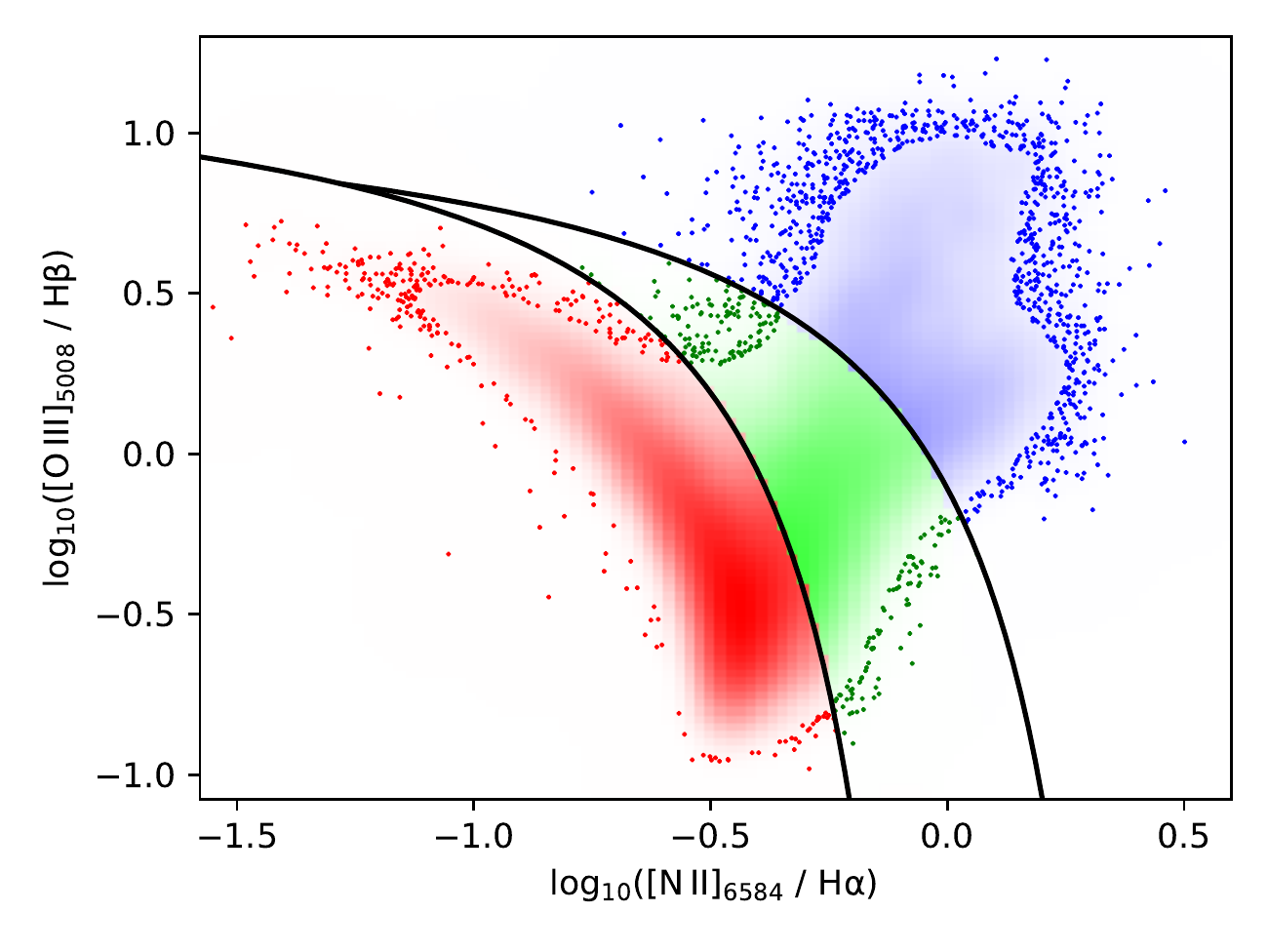}
\caption{Distribution of high S/N SDSS DR14 galaxies spectra in the BPT diagram, following the criteria defined in \protect\cite{Brinchmann2004}. The star-forming (SF) galaxies, AGN, and composite galaxies are plotted in red, blue, and green, respectively. The SF galaxies are free from AGN contamination, and are located at the bottom left of the diagram as the forbidden lines are weaker compared to the Balmer lines.}
\label{fig:BPT}
\end{figure}

We removed potentially problematic spectra from our sample by requiring the bitmask of warning \texttt{zWarning} to be zero, and the median signal-to-noise ratio in the r-band \texttt{snMedian} to be greater than 10. We also impose a constraint in redshift, $0.05<z<0.1$, to reduce systematics from redshift-induced trends (see Section~\ref{sec:z}). At these redshifts, the 3$^{\prime\prime}$ diameter of the fibres span a physical distance of 2.9 and 5.5 kpc, respectively. We selected spectra that are free from AGN contamination by using the BPT diagram \citep{Baldwin1981}, following the criteria defined in \cite{Brinchmann2004}. Galaxies with S/N$>$3 in H$\beta$, [\textsc{O\,iii}] $\lambda$5007, H$\alpha$ and [\textsc{N\,ii}] $\lambda$6584 are plotted in Figure~\ref{fig:BPT}. These galaxies are classified as star-forming (SF), AGN, or a mix of both (Composite) depending on their location on the BPT diagram. We chose only the unambiguously star-forming galaxies ($\texttt{bptclass}=1$), leaving us with 53283 galaxies in total.

\subsection{Sampling by galactic properties}
\label{sec:vdisp_sfr}

We sampled the spectra according to their stellar velocity dispersion ($v_{\rm disp}$) and star formation rate (SFR), which are the fundamental properties for our purposes. $v_{\rm disp}$ is most tightly correlated to the gravitational potential well of the galaxy and therefore the emission line width ($\sigma$), and was chosen to avoid superposition of emission lines with different widths during the stacking procedure. On the other hand, the SFR is directly correlated to the luminosity of emission lines in the absence of AGN contamination. We take the $v_{\rm disp}$ measurements directly from the SDSS \textsc{SpecObj} catalogue. The velocity dispersion estimates of the SDSS I-II spectra are obtained by directly fitting a set of stellar templates that match the resolution and sampling of the data, after being convolved with a gaussian kernel whose width is left as a free parameter. These estimates are deemed reliable at $\rm S/N>10$ and above 70 km s$^{-1}$, i.e. within our selection criteria\footnote{\url{https://classic.sdss.org/dr7/algorithms/veldisp.html}}. The SFR measurements are applied directly from the MPA-JHU catalogue\footnote{\url{https://www.sdss.org/dr14/spectro/galaxy_mpajhu/}} (\texttt{sfr\_fib\_p50}) that follow the prescription from \cite{Brinchmann2004}, using the H$\alpha$ luminosity corrected for dust attenuation. We did not apply any aperture correction, as the gaseous kinematics are inferred exclusively from the information within the fibre aperture.

\begin{figure}
\includegraphics[width=\columnwidth]{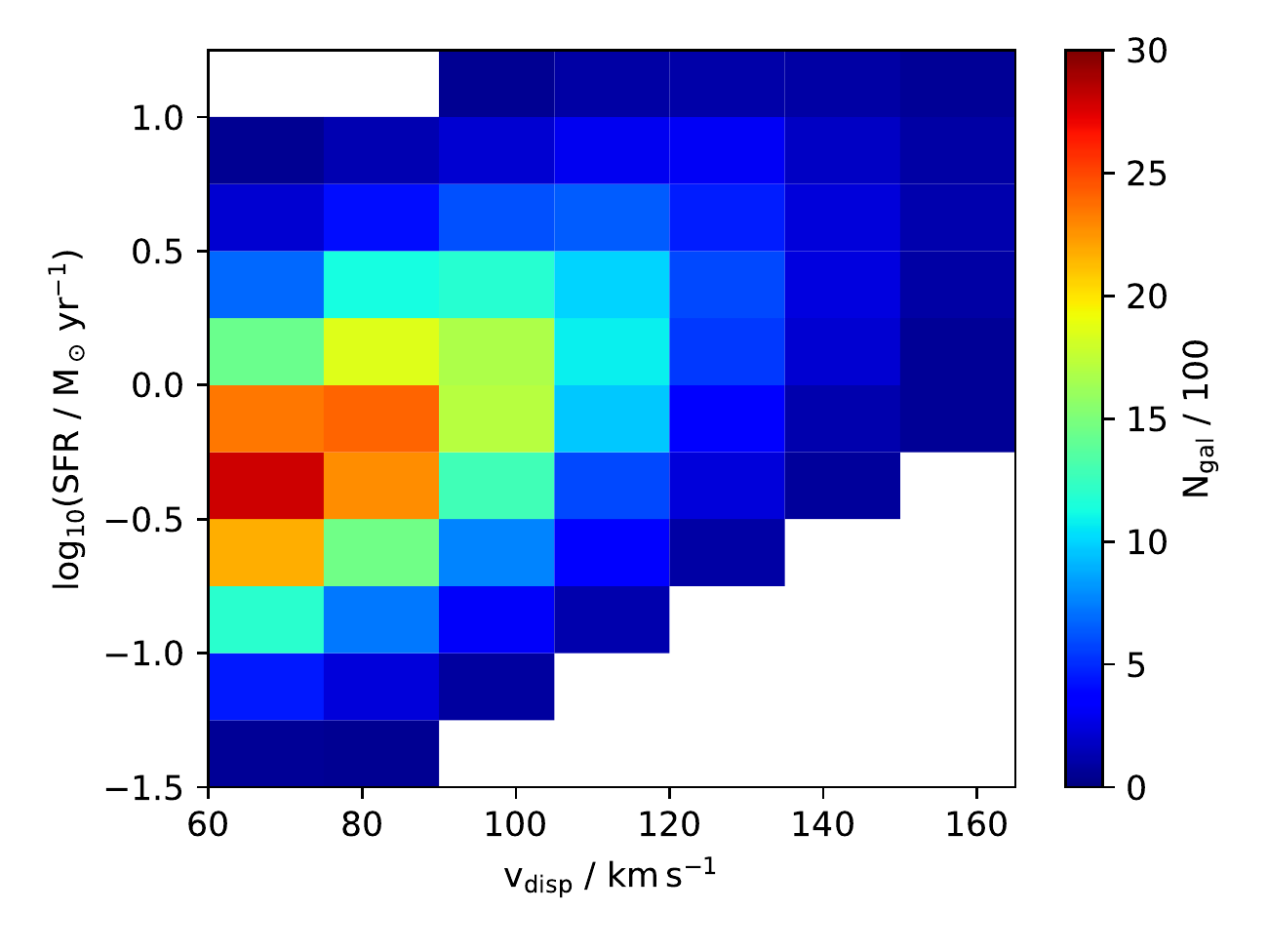}
\caption{Distribution of star-forming galaxies on the $v_{\rm disp}$-SFR plane. The plane is equally divided into fine grids, from $v_{\rm disp}=60$ km/s to 165 km/s with increments of 15 km/s, and from $\log_{10}({\rm SFR}\ /\ M_\odot\rm\,yr^{-1})=-1.5$ to 1.25 with increments of 0.25. A minimum of $N_{\rm gal}=50$ galaxy spectra is required for each bin, and the grid is colour-coded according to $N_{\rm gal}$.}
\label{fig:param_sfr}
\end{figure}

The distribution of galaxies on the $v_{\rm disp}$-SFR plane is plotted in Figure~\ref{fig:param_sfr}. The $v_{\rm disp}$-SFR plane was equally divided into fine grids, from $v_{\rm disp}=60~{\rm km\ s}^{-1}$ to $165~{\rm km\ s}^{-1}$ with increments of $15~{\rm km\ s}^{-1}$, and from $\log_{10}({\rm SFR}\ /\ M_\odot\rm\,yr^{-1})=-1.5$ to 1.25 with increments of 0.25. Each bin in the grid groups together galaxies with similar $v_{\rm disp}$ and SFR, and a minimum of 50 was required in each group to ensure the quality of the stacked spectra, resulting in 60 different groups across the parameter space according to Figure~\ref{fig:param_sfr}. This allows us to examine how the properties of emission lines vary across different values of $v_{\rm disp}$ and SFR in Section~\ref{sec:SFR}.

To examine how the line properties change with respect to relevant  galactic observables, we further split each group of galaxies into two subgroups according to the median parameter value. This method gives us limited number of subgroups, but effectively minimises the interdependence between our major parameters ($v_{\rm disp}$ and SFR) and other parameters. For example, if we stacked the galaxies according to their $v_{\rm disp}$ and specific SFR (sSFR), then we could not examine how the line properties vary across different sSFR irrespective of the SFR, as SFR and sSFR are directly correlated with each other.

We analysed the dependence of the line profiles on the axial ratio ($b/a$), 4000 \AA\ break strength ($D_n(4000)$, adopting the definition of \citealt{Balogh1999}), and sSFR in Sections \ref{sec:expAB}, \ref{sec:d4k}, and \ref{sec:ssfr}, respectively. We applied the sSFR measurements directly from the MPA-JHU catalog (\texttt{ssfr\_fib\_p50}), and the $b/a$ measurements directly from the SDSS \textsc{PhotoObjAll} catalog\footnote{\url{https://www.sdss.org/dr14/algorithms/classify/}} (\texttt{expAB\_r}).

\section{DATA PROCESSING}
\label{sec:process}

\subsection{Stacking procedure}
\label{sec:stack}

\begin{figure*}
\includegraphics[width=\textwidth]{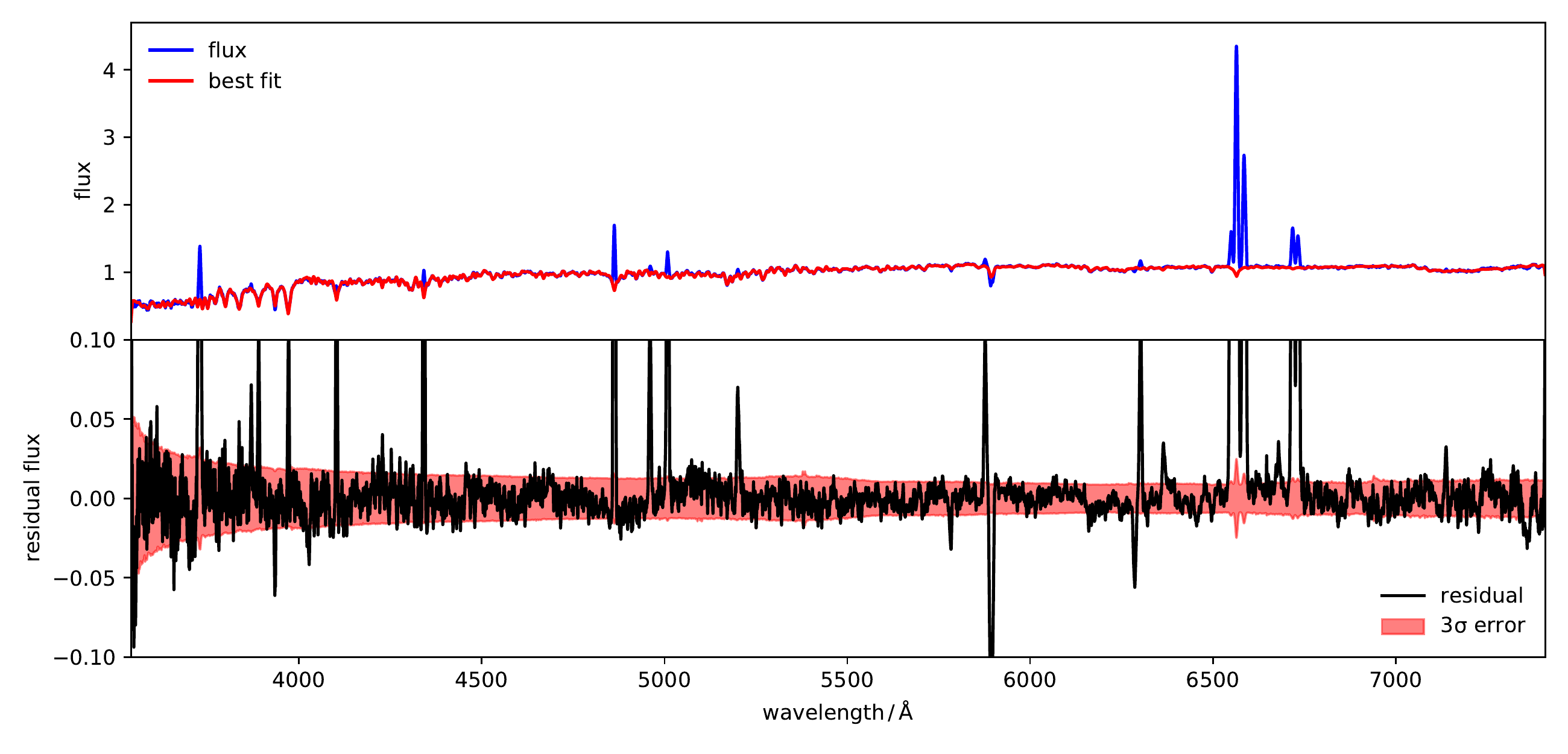}
\caption{Example of stellar continuum fitting using the pPXF algorithm over the wavelength range $3540.5\,\textup{\AA}\le\lambda\le7409.6\,\textup{\AA}$. The stacked spectrum is chosen from the $v_{\rm disp}$-SFR plane at $135\,{\rm km/s}\le v_{\rm disp}\le150\,{\rm km/s}$ and $0.75\le{\rm log_{10}(SFR}/M_\odot{\rm yr}^{-1})\le1$. The spectral flux and the continuum fit are plotted in blue and red respectively in the upper panel, and the residual is plotted in black in the lower panel. The error of the flux at 3$\sigma$ level (shaded in red) indicates that the stacked spectrum and the fit are in excellent agreement over the entire wavelength range. A detailed view of the H$\beta$ emission line fit for this spectrum is shown in Fig.~\ref{fig:kurt}.}
\label{fig:ppxf}
\end{figure*}

We combined the individual galaxy spectra to produce the stacked spectra that have significantly improved S/N, and follow a statistical approach instead of targeting individual cases. Since we are looking for departures from a standard Gaussian line profile, the stacking procedure maximises the detection of this trend. In particular, we probe a well-defined parameter space (see Figure~\ref{fig:param_sfr}) by stacking all spectra from the designated grid. The stacking procedure removes galaxy-to-galaxy variations, keeping the common properties of emission lines within the chosen grid. Before stacking, every galaxy spectrum was dereddened, deredshifted, and then renormalised to the median of its continuum flux between 5000 \AA{} and 5500 \AA{}. During the dereddening process, we took the g-band extinction coefficient $A_g$ directly from the SDSS photometric catalog (\texttt{extinction\_g}), converted it to $E(B-V)$ by applying the conversion factor $E(B-V)/A_g=3.793$ from \cite{Stoughton2002}, and used the extinction law from \cite{Cardelli1989}. We then shifted the spectrum back to its rest frame, and masked out all the problematic pixels within the spectrum by requiring the \texttt{AND} bitmask to be zero.

We used a variation of the Drizzle algorithm \citep{Fruchter2002}, a linear reconstruction method for undersampled images, to remap the galaxy spectra to a stacked spectrum \citep[see][]{Ferreras2013}. We set the sampling of the stacked spectrum to $\Delta\log_{10}(\lambda/$\AA$)=10^{-4}$ (roughly corresponding to $\Delta\lambda\sim$1\,\AA\ within the region of interest), the same as those of the SDSS spectra to maximise its S/N. To interpolate between adjacent spectral points, we did not split the stacked spectrum into finer bins during the stacking procedure in order to avoid diluting the spectral signal. Instead, we kept the same sampling and shifted the spectral points to the designated wavelengths and redid the stacking procedure. The maximum sampling $\Delta_{\rm max}$ of interpolated spectral points is determined by the precision of the redshift measurement from the MPA-JHU catalog, where $\Delta_{\rm max}=0.434\,\Delta z/(1+z)$. We found from our galaxy sample that the 5th and 95th percentiles of $\Delta_{\rm max}$ are $2.6\times10^{-6}$ and $1.2\times10^{-5}$ respectively, and so we set $\Delta_{\rm max}=10^{-5}$ for all interpolated spectral points \citep[see also][where $\Delta z\sim10^{-6}$]{Ferreras2016}. $\Delta z$ is substantially lower than the estimated outflow velocities that are determined in this work (see Section~\ref{sec:vout}, where $v_{\rm out}\approx 150~{\rm km\ s}^{-1}$, which is much greater than $\Delta_{\rm max}=6.9~{\rm km\ s}^{-1}$), so it is unlikely to introduce error in our measurements. This is a novel approach to interpolate the stacked spectra without diluting the signal, and it helps to better probe the kurtosis of the emission lines.

The RMS error of the flux and the RMS spectral resolution of the pixel were stacked in the same way. We estimated the error of the results using a bootstrapping technique, which will be discussed in detail in Section~\ref{sec:result}.

\subsection{Continuum fitting}
\label{sec:fit}

We removed the stellar continuum from the stacked spectra using the penalized pixel-fitting (pPXF) algorithm developed by \cite{Cappellari2004}, where the stellar continuum is seen as a linear combination of single stellar populations (SSP) convolved by the stellar line-of-sight velocity distribution (LoSVD). Both the SSP weights and LoSVD are optimised during the fitting process. We carried out the pPXF analysis on the stacked spectra over the wavelength range $3540.5\le \lambda/\textup{\AA}\le 7409.6$ to match the spectral templates from the stellar library, where we selected the MILES-based model of \cite{Vazdekis2010}, with FWHM resolution of 2.3 \AA{}. The ages and metallicities of the SSP range from 0.06 to 18 Gyr and from $10^{-2.32}$ to $10^{0.22}$ Z$_\odot{}$, respectively. During the stacking procedure, the gas emission lines were masked to suppress the contamination from non-stellar features. The spectral templates were convolved with the quadratic difference between the SDSS and the MILES instrumental resolution, which were then interpolated to the spectral points of the stacked spectra.

An example of the continuum fitting result is shown in Figure~\ref{fig:ppxf}. The stacked spectrum and the continuum fit are shown in blue and red respectively in the upper panel, and the residual is shown in black in the lower panel. The red shaded area denotes the RMS error of the flux to a 3$\sigma$ level, and shows an excellent agreement between the stacked spectrum and the fit over the entire wavelength range. Therefore, we can reliably use the residual spectra to analyse the emission features of the ionised gas. We also note that potential systematics from the residual noise on the characterisation of the emission lines will be mitigated by comparing a range of emission lines in different parts of the spectrum, thus affected by different regions of the stellar component.

\subsection{Profile model of emission lines}
\label{sec:kurt}

In the first order, an emission line can be crudely characterised by a Gaussian profile, where its line width is controlled by the LoSVD of the ionised gas throughout the galaxy. To address the subtle traits of emission lines, \cite{Cicone2016} fitted the emission line with a sum of three Gaussians. \cite{Concas2017} fitted the [\textsc{O\,iii}] $\lambda$5007 line from active galaxies with a double Gaussian to account for the AGN contribution towards the broader component and the wings of the line. \cite{Chen2016} fitted the H$\alpha$ line with a double Gaussian to calculate its kurtosis. Here, our focus lies on the emission lines of star-forming galaxies and measure their kurtosis and skewness. We consider a physically motivated model by accounting for the radially outward motion of the ionised gas in the presence of starburst-driven outflows, to explain the kurtosis of emission lines. We introduce a second Gaussian line component to the model if the emission lines are skewed.

The stacked spectra include a large number of star-forming galaxies that are free from AGN contamination with similar stellar velocity dispersion and SFR, but with various shapes and orientations. Therefore, a stacked spectrum represents a combined galaxy with an approximately spherical distribution of stars, gas and star-formation activity. This is significantly different from AGN galaxies, so we do not necessarily expect a broad, asymmetric component in the emission lines. Most of the star-formation activity happens at the galactic center, potentially driving a galactic-scale outflow radially outwards. We emulate the effect of outflow by Doppler shifting all the ionised gas radially at the same velocity $v_{\rm out}$. Along the line of sight, the projected Doppler velocity is equivalent to $v=v_{\rm out}\cos{\theta}$. Integrating on the sphere the projections of the velocity vector along the line of sight, 
we produce the corresponding Doppler profile\footnote{Note that this profile is valid when self-absorption of the line is insignificant. This assumption is adequate for this study, given that the radiative transfer effect caused by the line opacity is not the most significant factor that gives rise to the fluctuations and uncertainties in the spectral analysis. },
\begin{equation}
\label{eq:Doppler}
D(v,v_{\rm out})=\frac{1}{2v_{\rm out}} \ ,
\end{equation} 
for $|v|\leq v_{\rm out}$. This kernel is a simple boxcar function defined within the velocity interval $[-v_{\rm out},+v_{\rm out}]$, where the gas approaches the observer when $v=-v_{\rm out}$ and recedes from the observer when $v=v_{\rm out}$. Because the integral over $v=\pm v_{\rm out}$ is normalised to unity, Doppler shifting the ionised gas radially at velocity $v_{\rm out}$ is equivalent to convolving the emission line profile with the distribution function $D(v,v_{\rm out})$. If we assume that the emission line is adequately characterised by a single Gaussian in the absence of outflows, then the emission line profile is computed as
\begin{equation}\label{eq:model}
\begin{split}
F(\lambda)&=\int_{-\infty}^{+\infty}{\rm d}v \ 
D(v,v_{\rm out})\ \frac{A}{\sqrt{2\pi}\;\! \sigma_{\rm g}} \,
\exp\left[{-\frac{(\lambda-\lambda_0)^2}{2\;\! {\sigma_{\rm g}}^2}}\right]  \\
&=\frac{1}{4\;\! \Delta\lambda}
\left[{\rm erf}\left(\frac{\lambda-\lambda_0+\Delta\lambda}{\sqrt{2}\;\! \sigma_{\rm g}}\right)
-{\rm erf}\left(\frac{\lambda-\lambda_0-\Delta\lambda}{\sqrt{2}\;\!\sigma_{\rm g}}\right)\right]\ ,
\end{split}
\end{equation}
where erf is the error function, $\Delta\lambda=v_{\rm out}\lambda_0/c$ is the Doppler shift and $c$ is the speed of light. In summary, the emission line profile from our model depends on the three Gaussian variables: line amplitude $A$, mean wavelength $\lambda_0$, and Gaussian linewidth $\sigma_{\rm g}$, as well as one extra variable $v_{\rm out}$ to characterise the Doppler broadening of the ionised gas caused by galactic outflows. Such a convolution will lower and flatten the central peak and shrink the wings to result in a negative kurtosis that depends on both $\sigma_{\rm g}$ and $v_{\rm out}$ (see Equation~\ref{eq:kurtosis}). The total linewidth after the Doppler broadening is
\begin{equation}
\label{eq:total_width}
\sigma=\sqrt{{\sigma_{\rm g}}^2+{\sigma_{\rm out}}^2+{\sigma_{\rm inst}}^2}\ ,
\end{equation}
where $\sigma_{\rm out}=\Delta\lambda/\sqrt{3}$. $\sigma_{\rm inst}$ represents the broadening of intrinsic emission lines due to the SDSS spectral resolution, as well as the convolution of bin width during the stacking process.

We determine the strength of the outflow from the kurtosis ($\kappa$) of the emission line. Since $\kappa_{\rm g}=0$ for the Gaussian distribution and $\kappa_{\rm out}=-1.2$ for $D(v,v_{\rm out})$, the overall kurtosis is:
\begin{equation}
\label{eq:kurtosis}
\kappa=\frac{{\sigma_{\rm g}}^4\kappa_{\rm g}+{\sigma_{\rm out}}^4\kappa_{\rm out}}{({\sigma_{\rm g}}^2+{\sigma_{\rm out}}^2)^2}=\frac{-1.2\,{\sigma_{\rm out}}^4}{\sigma^4}\ ,
\end{equation}
where $\kappa$ decreases as $v_{\rm out}$ increases. Note that we adopt the definition of {\sl excess kurtosis} for all measurements of this parameter, and the instrumental effect is assumed to be negligible, i.e. $\kappa_{\rm inst}=0$ \citep[see, e.g.][]{Law2021}. Such assumption is also justified by the presence of Gaussian line profiles from high S/N stacked spectra of galaxies at low SFR as shown in Section~\ref{sec:SFR}.

Our emission line fitting pipeline comprises four stages: 1) the line is fitted with a single Gaussian; 2) the line is fitted with one convolved line $F(\lambda)$; 3) the line is fitted with two Gaussians; finally, 4) the line is fitted with one convolved line $F(\lambda)$ and an additional Gaussian. In each stage, the goodness of fit is determined by the reduced chi-squared statistic, $\chi^2$. The fit was considered adequate if $\chi^2\le1.5$, and the later stages of the fit were not required. If the convolved line profile $F(\lambda)$ was not needed, the kurtosis is considered to be zero. Likewise, if the second Gaussian component was not needed, the skewness is considered to be zero. When the second Gaussian component is present, the skewness is computed analytically as
\begin{equation}
s=\frac{\sum_{i=1}^{2}A_i\left[3{\sigma_i}^2 \left(\lambda_{0,i}-\bar{\lambda}_0\right)+\left(\lambda_{0,i}-\bar{\lambda}_0\right)^3\right]}{\left[\sum_{i=1}^{2}A_i\left({\sigma_i}^2+\left(\lambda_{0,i}-\bar{\lambda}_0\right)^2\right)\right]^{3/2}}
\end{equation}
where $\bar{\lambda}_0$ is the mean wavelength of the overall emission line.

\begin{figure}
\includegraphics[width=\columnwidth]{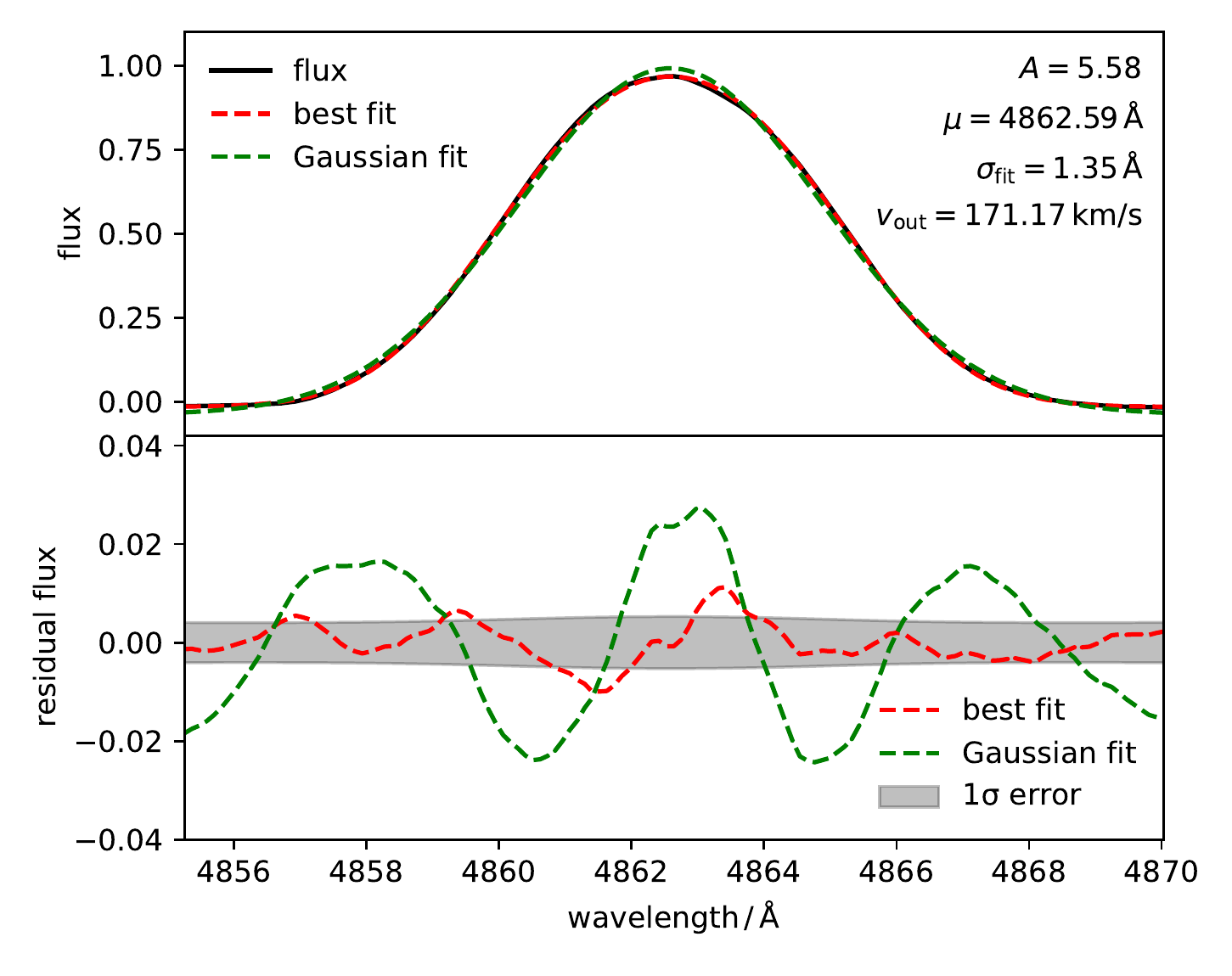}
\caption{Example of H$\beta$ emission line fitted with our line model, $F(\lambda$), which accounts for the presence of an outflow. The emission line and our best fit are plotted in solid black and dashed red respectively, complemented by the single Gaussian fit in dashed green. The flux is plotted in the upper panel, and the residual flux is plotted in the lower panel which also includes the error of the flux at 1$\sigma$ level. The Gaussian fit is inadequate as the emission line is platykurtic, where deficit is found at the wings and the peak of the line, and excess is found between the wings and the peak. The H$\beta$ line is well fitted by our model (i.e., a second component is not needed), and the parameters are shown in the top right corner of the upper panel.}
\label{fig:kurt}
\end{figure}

As an example, we fit the H$\beta$ emission line from Figure~\ref{fig:ppxf} and show the result in Figure~\ref{fig:kurt} to demonstrate that the H$\beta$ line is well fitted by our emission line model (i.e., a second component is not needed). We complement our model with a single Gaussian fit, showing that it is inadequate as the emission line is clearly platykurtic ($\kappa<0$), where deficit is found at the wings and the peak of the line, and excess is found between the wings and the peak. In the bottom panel of Figure~\ref{fig:kurt}, we compare the residual of our best fit (in red) that assumes non-zero kurtosis with that of a Gaussian fit (in green). The actual uncertainty of the spectra is shown as a grey shaded region, which confirms the non-Gaussianity of the line. Note that the measured kurtosis is consistent in various emission lines across the spectrum (see Section~\ref{sec:SFR}), which shows that the measurements are resilient to any potential residual and inaccuracy due to the stellar population fit. In addition, the results of stacked galaxies are consistent with those of individual galaxies (see Section~\ref{sec:individual}), which shows that the stacking procedure is robust and does not affect the measurements of kurtosis. More examples are provided in \ref{sec:app1} to justify that the platykurtic line profile cannot be simply due to residuals and inaccuracies in the procedures of data processing.

\section{RESULTS}
\label{sec:result}

Our analysis is applied to the most prominent emission lines in star-forming systems. For clarity, we first present the results for the H$\beta$, [\textsc{O\,iii}] $\lambda$5008, [\textsc{N\,ii}] $\lambda\lambda$6549,6564, H$\alpha$, and [\textsc{S\,ii}] $\lambda\lambda$6718,6732 emission lines in Section~\ref{sec:SFR}. We show that the non-Gaussian features are present in all lines considered, except for [\textsc{O\,iii}]. Then, we choose H$\beta$ as a representative emission line due to the absence of contamination from adjacent lines, to avoid presenting redundant results when analysing the dependence of line properties on relevant galactic observables in later subsections.

\begin{figure*}
\includegraphics[width=.98\textwidth]{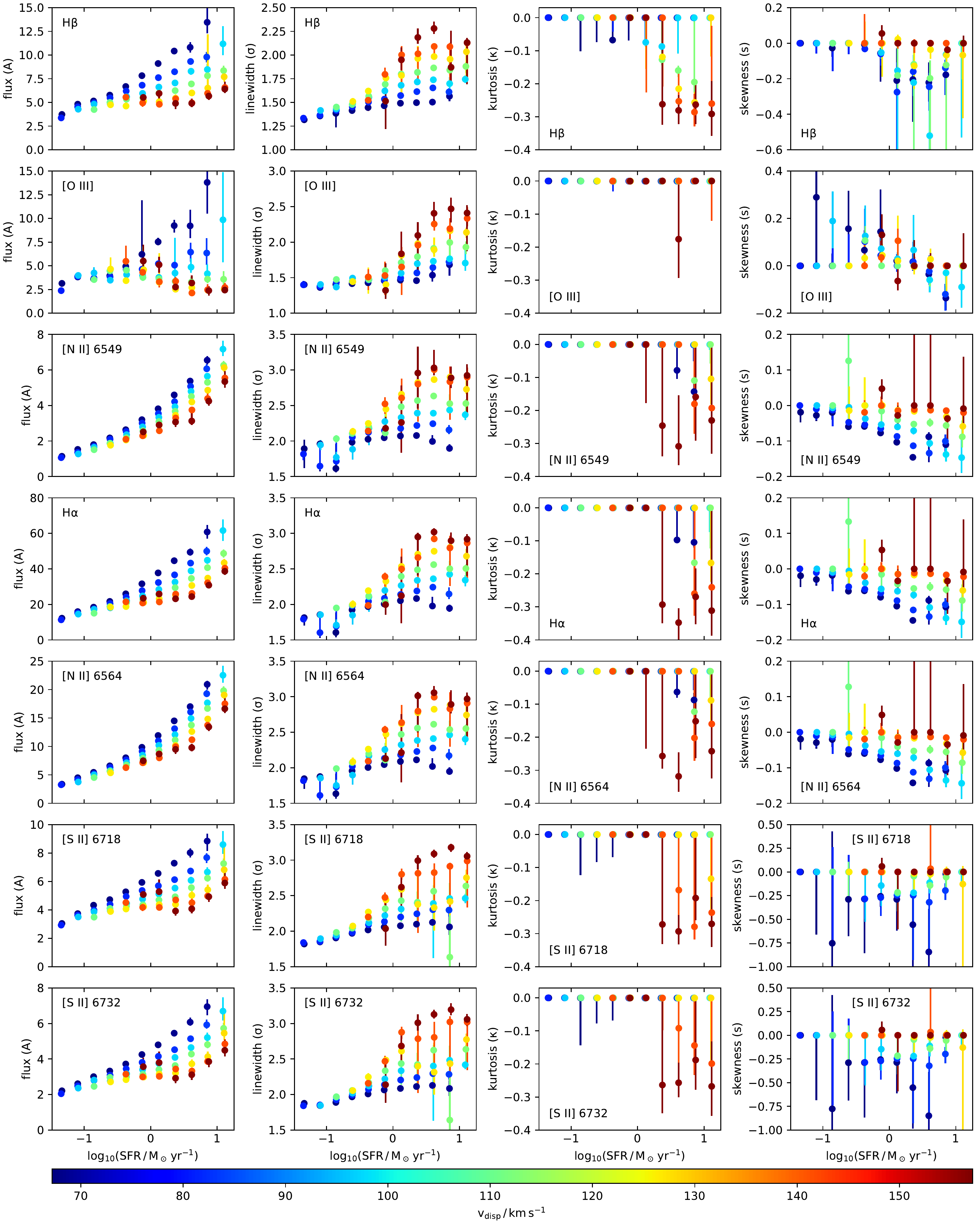}
\caption{Properties of various emission lines across the parameter space in Figure~\ref{fig:param_sfr}, where the scatter points are colour-coded based on the $v_{\rm disp}$ value. The first, second, third, and fourth column show the line amplitude $A$, linewidth $\sigma$, kurtosis $\kappa$, and skewness $s$, respectively. All panels share the same x-axis (SFR). The emission lines in strong star-forming systems feature negative kurtosis (with [\textsc{O\,iii}] being an exception), which show that the gas is radially accelerated according to our model in Section~\ref{sec:kurt}. Note that the error bars are estimated using a bootstrapping technique.}
\label{fig:sfr}
\end{figure*}

\subsection{Star formation rate}
\label{sec:SFR}

The main result is shown in Figure~\ref{fig:sfr}, where the flux ($A$), linewidth ($\sigma$), kurtosis ($\kappa$), and skewness ($s$) are plotted in the first, second, third, and fourth column, respectively. The colour of the scatter points represents the stellar velocity dispersion ($v_{\rm disp}$) of the stacked galaxies, which is proportional to their gravitational potential well, and therefore, to their mass. High and low mass galaxies are respectively distinguished by the red and blue ends of the colour map.

The line amplitude ($A$) panels show that the flux of the emission lines is directly proportional to the SFR for the most part, which is to be expected because the SFR estimation from the MPA-JHU catalogue adopted the method from \cite{Brinchmann2004}. However, the line amplitude of [\textsc{O\,iii}] is inversely proportional to the SFR for massive galaxies, and such exception suggests that the [\textsc{O\,iii}] line traces a different gas component. At low SFR, the line amplitude flattens out towards the faint end, which implies that the strength of the absorption lines may have been underestimated when fitting the stellar continuum of individual low S/N galaxies at low SFR. Note that $A$ is renormalised with respect to the continuum flux between 5000 \AA{} and 5500 \AA{}, meaning that the equivalent widths of emission lines are higher at lower $v_{\rm disp}$ in general.

The linewidth ($\sigma$) panels show that for all emission lines, $\sigma$ increases with increasing $v_{\rm disp}$ and SFR. This result is consistent with the finding in \cite{Cicone2016}, where the emission lines are broadened by galactic outflows. Moreover, at fixed $v_{\rm disp}$, $\sigma$ increases slowly with increasing SFR at low $v_{\rm disp}$, but increases more rapidly with increasing SFR at high $v_{\rm disp}$ and flattens out at $\log_{10}({\rm SFR}\ /\ M_\odot\rm\,yr^{-1})\sim0.5$.

The kurtosis ($\kappa$) panels show that most emission lines are platykurtic ($\kappa<0$), except for the [\textsc{O\,iii}] line which is mostly mesokurtic ($\kappa=0$). In our model (see Section~\ref{sec:kurt}), a negative kurtosis in the emission line implies that the kinematics of the line emitting gas is directly influenced by starburst-driven galactic outflows, causing the gas to be radially accelerated. Such effect is the strongest for massive galaxies at high SFR. However, the [\textsc{O\,iii}] emission line is Gaussian-like, which suggests that the [\textsc{O\,iii}] line traces a different gas component and is excited turbulently by galactic outflows.

The skewness ($s$) panels show that the emission lines are negatively skewed for the most part. We attribute the negative skewness to obscuration by dust in the galactic disk \citep{Villar-Martin2011, Soto2012, Cicone2016}. The backside receding gas is more severely affected by dust extinction, thereby suppressing the red wing of the emission line. The skewness signature is the strongest for small galaxies at high SFR, as the amount of dusty content in the galactic disk is proportional to the SFR \citep{DaCunha2010, Hjorth2014}. Again, the [\textsc{O\,iii}] line is different from the rest and is overall more positively skewed.

Note that forbidden lines such as [\textsc{S\,ii}] are enhanced in shocks, expected in an outflow environment, in contrast with the [\textsc{O\,iii}] $\lambda$5007 line, which requires a hard ionizing radiation field to be enhanced in outflows. Our results in the analysis are consistent across the H$\beta$, [\textsc{N\,ii}], H$\alpha$, and [\textsc{S\,ii}] lines, which show that they are unaffected by any potential bias from the wavelength-dependent spectral resolution of the data.

\subsection{Axial ratio of galaxies}
\label{sec:expAB}

\begin{figure*}
\includegraphics[width=\textwidth]{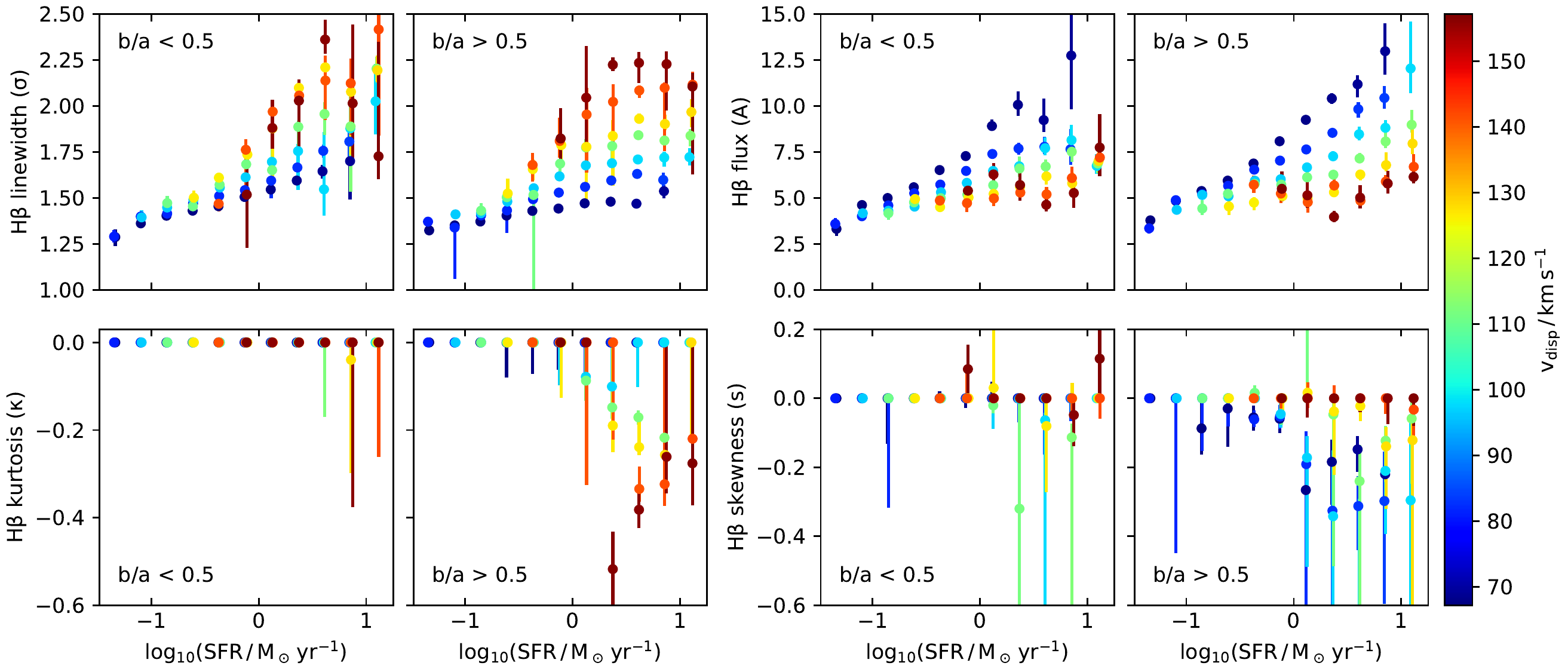}
\caption{Line properties of H$\beta$ from edge-on ($b/a<0.5$) and face-on ($b/a>0.5$) galaxies. Each group of galaxies from the same grid in Figure~\ref{fig:param_sfr} was separated into two subgroups basing on their axial ratio $b/a$, which are paired and compared. The linewidth $\sigma$, line amplitude $A$, kurtosis $\kappa$, and skewness $s$ are plotted in the top-left, top-right, bottom-left, and bottom-right panels respectively, where the scatter points are colour-coded based on the $v_{\rm disp}$ value. The line properties such as negative kurtosis and skewness that were found in Figure~\ref{fig:sfr} are amplified for face-on galaxies and suppressed for edge-on galaxies. This supports our hypothesis that the negative kurtosis is driven by galactic outflows which accelerates the gas radially outward, and can influence the line shape only if the outflowing gas is accelerated in the line of sight, i.e. preferably in face-on galaxies.}
\label{fig:expAB}
\end{figure*}

The H$\beta$ lines from massive galaxies with high SFR are found to be platykurtic, and those from small galaxies with high SFR are negatively skewed. We propose that the negative kurtosis is caused by galactic outflows which accelerate the gas radially outward, and the negative skewness is caused by dust obscuration of the galactic disk which affects the backside receding gas more severely. To verify this, we split each group of galaxies into two subgroups according to their axial ratio ($b/a$), a proxy to split the sample into edge-on galaxies ($b/a<0.5$) and face-on galaxies ($b/a>0.5$). If the negative kurtosis and skewness are caused by galactic outflows and star formation, such line properties should be more prominent in face-on galaxies, where outflows are accelerated mostly towards the line of sight and can be obscured by the galactic disk. The line properties of face-on and edge-on galaxies are shown in Figure~\ref{fig:expAB}, where the negative kurtosis and skewness are magnified for face-on galaxies and suppressed for edge-on galaxies. This supports our hypothesis that these line properties are driven by galactic outflows and star formation. In addition, the signatures of negative kurtosis and negative skewness are almost completely absent in edge-on galaxies. This suggests that the outflows are driven along the galactic rotation axis, presumably the path of least resistance, with a half angle that is smaller than $\theta<\cos^{-1}{0.5}$. This will be discussed further in Section~\ref{sec:vout}.

\subsection{4000 \AA{} break}
\label{sec:d4k}

\begin{figure*}
\includegraphics[width=\textwidth]{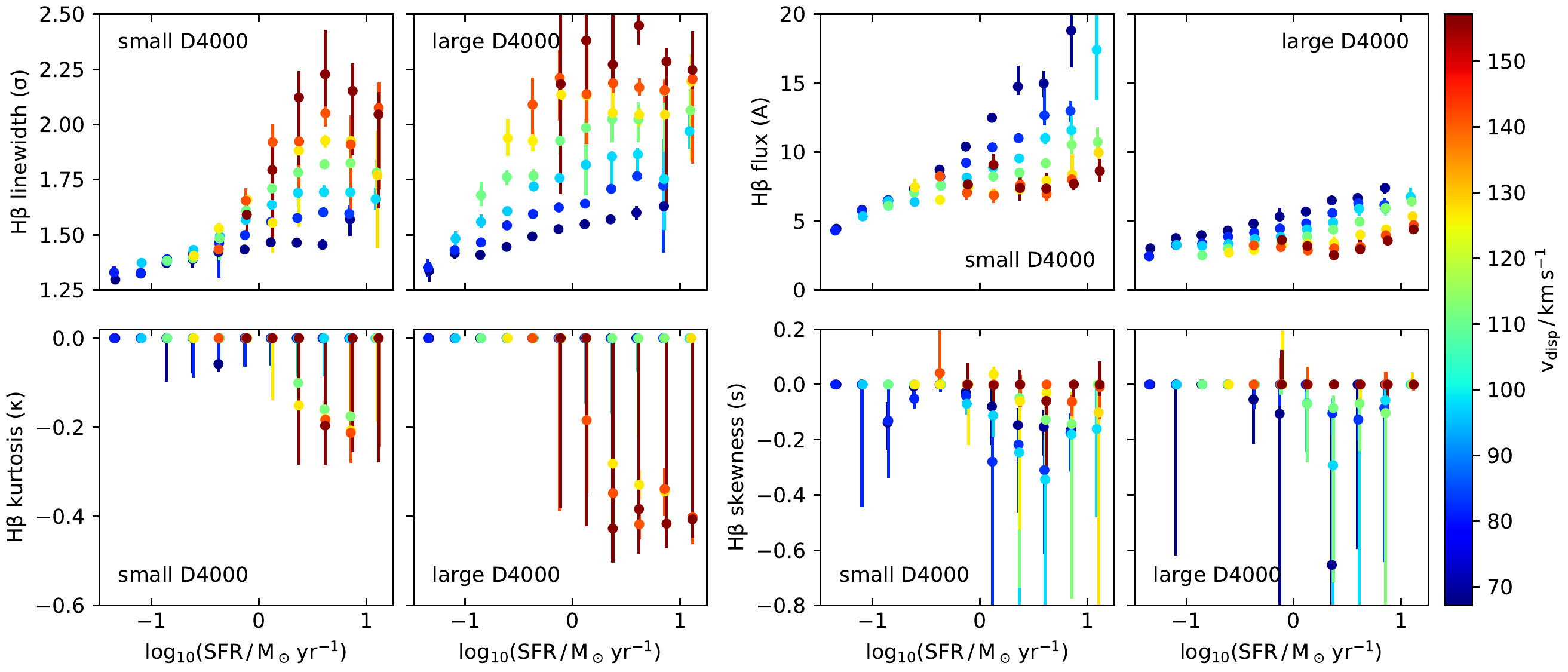}
\caption{Line properties of H$\beta$ from younger (small $D_n(4000)$) and older (large $D_n(4000)$) galaxies. Each group of galaxies from the same grid in Figure~\ref{fig:param_sfr} was separated into two subgroups based on its median $D_n(4000)$ value, which are paired and compared. The linewidth $\sigma$, line amplitude $A$, kurtosis $\kappa$, and skewness $s$ are plotted in the top-left, top-right, bottom-left, and bottom-right panels respectively, where the scatter points are colour-coded based on the $v_{\rm disp}$ value. The H$\beta$ lines are more platykurtic for older galaxies, which shows that the impact of galactic outflows can be increased and accumulated for galaxies which sustained a longer period of star formation. $\sigma$ at low SFR is significantly larger in older galaxies, which indicates that they have undergone recent starburst with imprint of star-formation from the previous starburst episode.}
\label{fig:d4k}
\end{figure*}

Complementary to the instantaneous SFR, whose estimate is based on the H$\alpha$ flux, tracing very young stellar populations (few tens of Myr), we target the 4000 \AA{} break (via the standard $D_n(4000)$ index) to measure the average stellar age of the galaxy. We examine how the line properties change according to $D_n(4000)$ by splitting each group of galaxies into two subgroups, with lower and higher $D_n(4000)$ values corresponding to younger and older galaxies, respectively\footnote{Note that for young stellar populations -- as expected in star-forming systems -- the age dependence of D$_n$(4000) dominates over the metallicity dependence \citep[e.g.][]{Bruzual2003}.}. This allows us to investigate not only the instantaneous effects of galactic outflows, but also how they impact the galaxies over a longer period of time. The line properties of younger and older galaxies are shown in Figure~\ref{fig:d4k}. The H$\beta$ lines from older galaxies are more platykurtic than those from younger galaxies, which implies that galactic outflows increase and accumulate their impact on the gas in older galaxies that sustained a longer period of star formation. Additionally, at low SFR, the linewidth ($\sigma$) of H$\beta$ in older galaxies is significantly larger than those in younger galaxies. This suggests that these older galaxies have undergone recent activity with imprint of star-formation (enhanced $\sigma$) from the previous starburst episode.

\subsection{Specific star formation rate (sSFR)}
\label{sec:ssfr}

\begin{figure*}
\includegraphics[width=\textwidth]{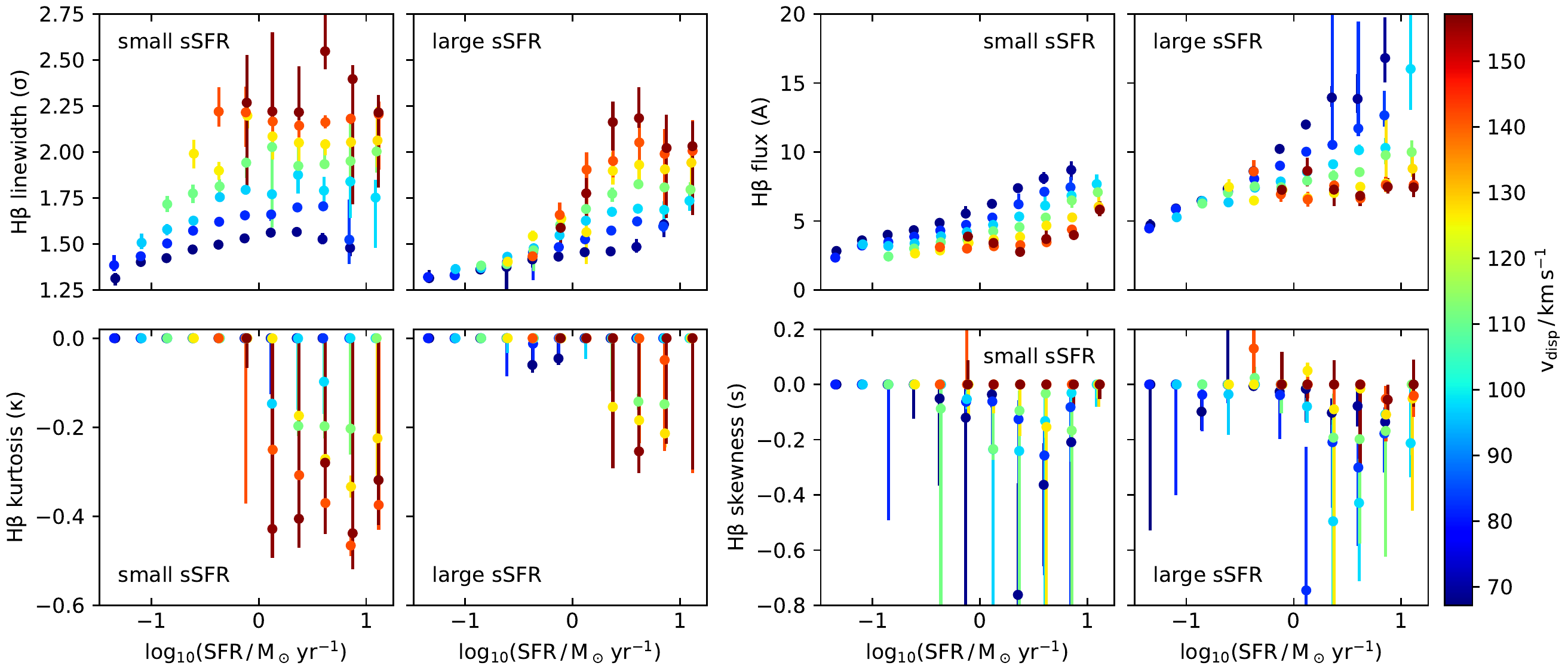}
\caption{Line properties of H$\beta$ from larger (small sSFR) and smaller (large sSFR) galaxies. Each group of galaxies from the same grid in Figure~\ref{fig:param_sfr} was separated into two subgroups based on its median sSFR value, which are paired and compared. The linewidth $\sigma$, line amplitude $A$, kurtosis $\kappa$, and skewness $s$ are plotted in the top-left, top-right, bottom-left, and bottom-right panels respectively, where the scatter points are colour-coded based on the $v_{\rm disp}$ value. The H$\beta$ lines are more platykurtic for larger galaxies, as they are in general older and have experienced a longer period of star formation.}
\label{fig:ssfr}
\end{figure*}

\begin{figure*}
\includegraphics[width=\textwidth]{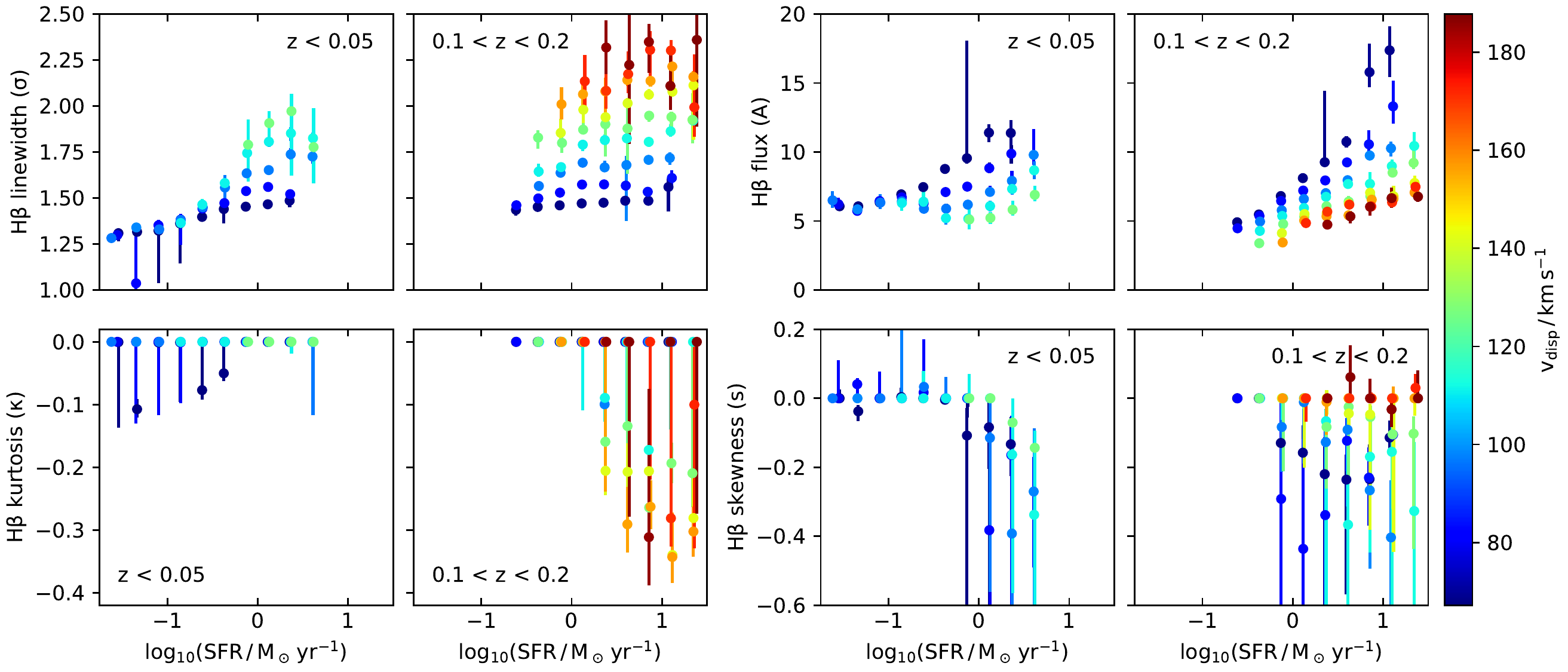}
\caption{Line properties of H$\beta$ from galaxies with lower redshift ($z<0.5$) and higher redshift ($0.1<z<0.2$). The linewidth $\sigma$, line amplitude $A$, kurtosis $\kappa$, and skewness $s$ are plotted in the top-left, top-right, bottom-left, and bottom-right panels respectively, where the scatter points are colour-coded based on the $v_{\rm disp}$ value. Since the SDSS spectra are taken within a fixed aperture, galaxies at lower and higher redshifts tend to be less massive with lower SFR and more massive with higher SFR, respectively. Overall, these results are consistent with our working sample, at redshift $0.05<z<0.1$ as shown in Figure~\ref{fig:sfr}.}
\label{fig:z}
\end{figure*}

In addition to $b/a$ and $D_n(4000)$, we also examine the effects of sSFR on the emission line properties. Each group of galaxies was split into two subgroups, with lower and higher sSFR values representing higher and lower mass galaxies (as they have similar SFR), respectively. If we assume that an increased density of SFR will enhance the power of galactic outflows and play a significant role in affecting the global gas kinematics, then we should expect the H$\beta$ line to be more platykurtic in smaller galaxies with higher sSFR. The line properties of larger and smaller galaxies are shown in Figure~\ref{fig:ssfr}, which does not favour the assumption and show that the H$\beta$ lines from more massive galaxies are actually more platykurtic. This is because sSFR is anti-correlated with $D_n(4000)$ (compare Figure~\ref{fig:d4k} to \ref{fig:ssfr}, their flux panels in particular), and larger galaxies are in general older and have experienced longer duration of star formation.

\section{DISCUSSION}
\label{sec:discuss}

\subsection{Redshift}
\label{sec:z}

In this study we have considered a sample of galaxies in a relatively narrow redshift range, $0.05<z<0.1$. This selection mitigates the aperture effect caused by using fibres with a fixed 3$^{\prime\prime}$ diameter. Here, we investigate the effect of choosing a wider redshift window for the selection. We study the H$\beta$ emission line from galaxies at lower redshift ($z<0.05$) and higher redshift ($0.1<z<0.2$). The results are shown in Figure~\ref{fig:z}, where we continue to see a platykurtic H$\beta$ line profile in more massive galaxies with higher SFR. Due to the flux limit of the survey, galaxies at lower and higher redshift tend to be less massive with lower SFR and more massive with higher SFR, respectively. Therefore, these two galaxy samples, to a large extent, can act as extensions to either side of the parameter space. These results are consistent with those from Figure~\ref{fig:sfr}: $\sigma$ depends primarily on $v_{\rm disp}$ and increases monotonically with SFR at low SFR (but flattens out at high SFR); the equivalent width of the H$\beta$ line decreases for more massive galaxies; the H$\beta$ line is the most platykurtic in massive galaxies with high SFR; the H$\beta$ line in less massive galaxies with high SFR is negatively skewed. Note that we are only interested here in confirming that the redshift of the sample does not introduce any significant systematic trend. While aperture bias will produce absolute differences, Figure~\ref{fig:z} confirms that the trends are unaffected. Since the results are consistent, we also conclude that aperture bias is minimal because the H$\beta$ intensity is significantly weaker at larger distances from the galactic centre.

\subsection{Comparison with previous work}
\label{sec:vout}

Signatures of outflows in SDSS galaxies were investigated in many occasions. \cite{Chen2010} analysed the Na\,I D absorption line in star-forming galaxies and found an outflow signature (where the gas component is blueshifted with respect to the stellar component) in face-on galaxies. This is consistent with our results in Section~\ref{sec:expAB} and Figure~\ref{fig:expAB}. They also found that the outflowing gas covers a wide angle, with strong outflow signature within $60^\circ$ from the disk rotation axis (equivalent to an opening angle of $120^\circ$). A galaxy inclination of $60^\circ$ is roughly equivalent to $b/a=0.5$ \citep[see][]{Padilla2008}. We have found a clear outflow signature is present only in galaxies with $b/a>0.5$, which is consistent with the opening angle found by \cite{Chen2010}.

\begin{figure}
\includegraphics[width=\columnwidth]{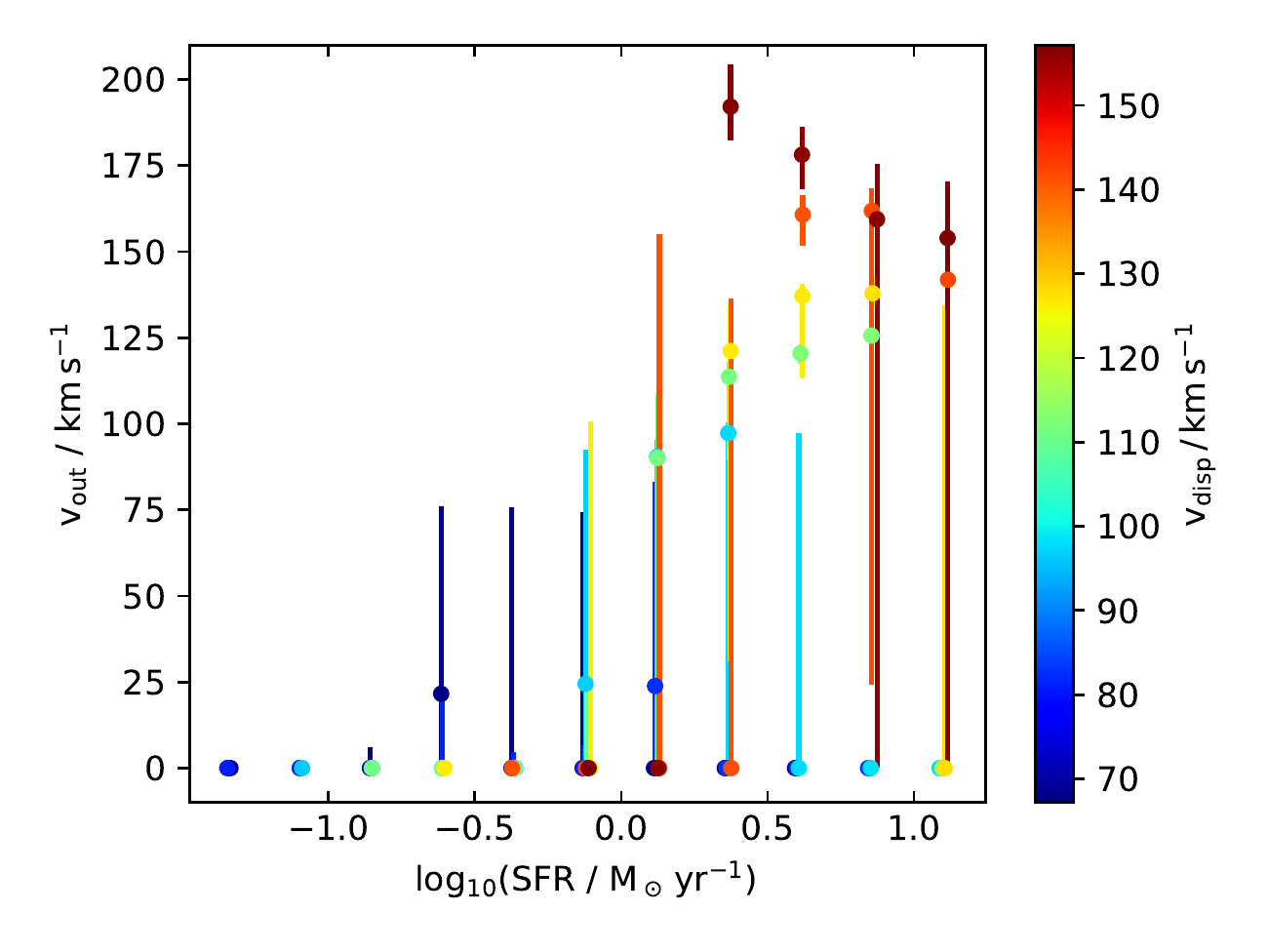}
\caption{Outflow velocity ($v_{\rm out}$) of galaxies with respect to star-formation rate (SFR, x-axis) and stellar velocity dispersion ($v_{\rm disp}$, colour-coded). Taking the kurtosis and linewidth of H$\beta$ emission line from face-on galaxies (as shown in Figure~\ref{fig:expAB}) into account, $v_{\rm out}$ is calculated by applying Equations \ref{eq:total_width} and \ref{eq:kurtosis}. $v_{\rm out}$ correlates strongly with $v_{\rm disp}$ but hardly depends SFR. Errors are estimated using a bootstrapping technique, where the scatter points represent the median and the bars represent the 1$\sigma$-confidence interval. If the error bar is large, the corresponding scatter point is always located on either side of the limit, implying that the transition is abrupt and $v_{\rm out}$ is bi-modal. This also applies to Figure~\ref{fig:sfr}$-$\ref{fig:z}.}
\label{fig:vout}
\end{figure}

To compare our results with the outflow velocities ($v_{\rm out}$) from the previous works, we applied Equations \ref{eq:total_width} and \ref{eq:kurtosis} to calculate $v_{\rm out}$ by using the kurtosis and linewidth of the H$\beta$ line in face-on galaxies (which can be found in Figure~\ref{fig:expAB}). The result is plotted in Figure~\ref{fig:vout}, where $v_{\rm out}$ correlates tightly with $v_{\rm disp}$ and is consistent with that of the Na\,I D results from \cite{Chen2010}, where $v_{\rm out}$ varied from $\sim120$ to $160\rm\ km\ s^{-1}$.

\citet{Cicone2016} analysed the H$\alpha$+[\textsc{N\,ii}] and [\textsc{O\,iii}] emission lines in star-forming galaxies and found that the outflow velocity of ionised gas correlates tightly with galactic stellar mass ($M_\star$). Since $M_\star$ is directly proportional to the square of ${v_{\rm disp}}$, we argue that this is consistent with our results. \cite{Cicone2016} also found that the outflow velocity correlates tightly with SFR for SFR$>$1\,M$_\odot$\,yr$^{-1}$, where the deviations in the gas kinematics from stellar kinematics increase with increasing SFR. Our results in Section~\ref{sec:SFR} and Figure~\ref{fig:sfr} shows that the gas kinematics, as traced by $\sigma$, correlates tightly with SFR at fixed $v_{\rm disp}$, in particular for ${\rm SFR}>1\ {\rm M}_\odot\rm\ yr^{-1}$. This implies that the gas and stellar kinematics are increasingly deviated at higher SFR, strengthening the outflow signature according to the definition of \cite{Cicone2016}. In addition,
they also detected the line asymmetry by measuring the difference between the 2.3th and 97.7th percentile velocities to show that the blue asymmetry (negative skewness) is present in almost all cases, which is consistent with our results in Section~\ref{sec:SFR}.

While the results of $v_{\rm out}$ are consistent, we have adopted a different interpretation of outflow signature, where we measure $v_{\rm out}$ by tracing the bulk movement of gas in the radial direction which can be quantified using the line kurtosis according to Section~\ref{sec:kurt}. We argue that the high-velocity tail of emission lines traces gas at super-virial temperature, which may not necessarily be equivalent to the typically bi-conical outflows driven by star-formation. For the same reason, \cite{Concas2017} argued that a second broader gas component must be present if the host galaxy is experiencing outflow. By analysing the [\textsc{O\,iii}] line profile, they found no significant evidence for outflows from star-forming galaxies. This is consistent with our results in Section~\ref{sec:SFR}, where the [\textsc{O\,iii}] emission line is well fitted by Gaussian lines, i.e. zero kurtosis. \cite{Concas2017} argued that the additional broad component can only be found in active galaxies and concluded that star formation does not play a primary role in driving galactic-scale outflows, which contradicted the results from \cite{Chen2010} and \cite{Cicone2016}. However, the interstellar and circumgalactic gas is a stratified multi-phase fluid which spans a wide range of physical conditions \citep{Tumlinson2017}, and the [\textsc{O\,iii}] forbidden line is expected to arise in an environment that is more diffuse than that of the Balmer lines \cite[see, e.g.,][]{Osterbrock2006}. Therefore, we believe that the [\textsc{O\,iii}] line traces a different component that is located in the outskirts of galaxies, external to the H$\beta$ emitting gas, resolving the seemingly contradictory conclusions from previous works. Note that the outflowing gas is also stratified \citep[e.g.][]{Chevalier1985}, causing the rate of interaction between the [\textsc{O\,iii}] emitting gas and the outflowing gas to be significantly lower, so that the signature of negative kurtosis is generally absent from the [\textsc{O\,iii}] emission line. Interestingly, if we extended our analysis to active galaxies, we would expect the line profile of [\textsc{O\,iii}] to be leptokurtic ($\kappa>0$) due to the additional broad component introduced by AGN activity.

\begin{figure*}
\includegraphics[width=\textwidth]{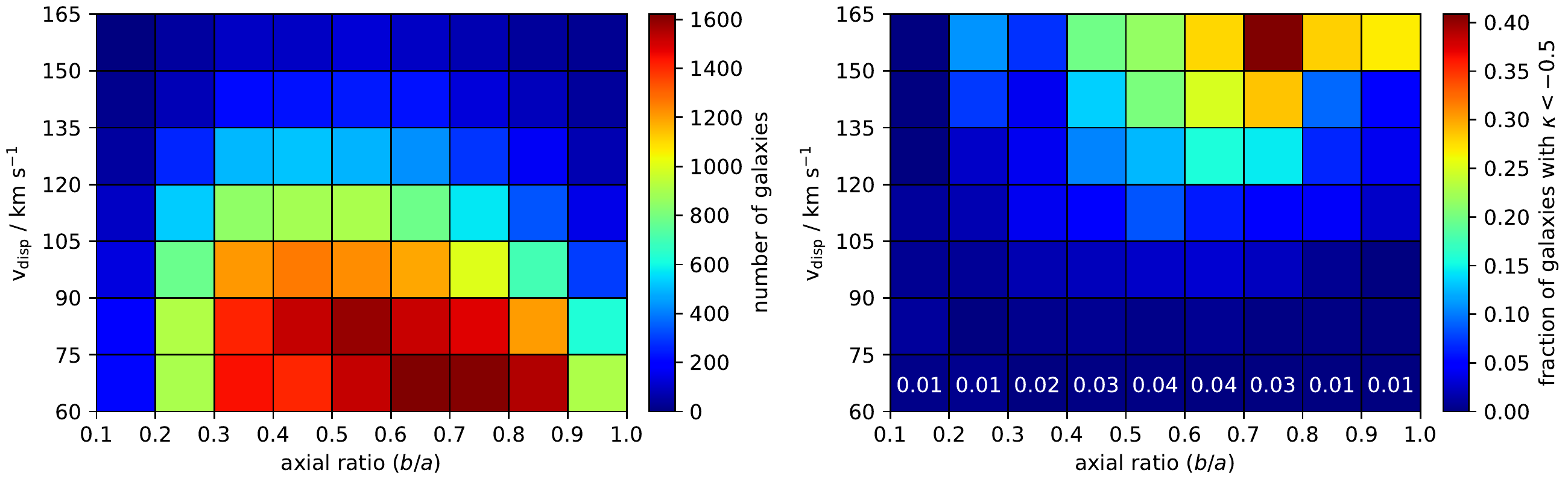}
\caption{Galaxy population distribution with respect to axial ratio ($b/a$) and stellar velocity dispersion ($v_{\rm disp}$) on the left panel, and the corresponding fraction, $F$ of galaxies with H$\beta$ kurtosis $\kappa<-0.5$ on the right panel. This figure is produced by applying the same methodology we have presented in the analysis of stacked spectra, but this time using the spectra of individual galaxies, from which the fraction $F$ is calculated. The color maps show that the platykurtic H$\beta$ line profile is more common in galaxies with larger $b/a$, i.e. face-on galaxies, consistent with our results in Section~\ref{sec:expAB}. The white numbers at the bottom of the right panel represent the average $F$ across the entire column, which are in first order consistent with the results from \protect\cite{Chen2016}. This shows that it is necessary to take into account that each $b/a$ column is mixed with galaxies of different $v_{\rm disp}$, as the average $F$ values are misrepresented by lower mass galaxies which dominate the overall population.}
\label{fig:boxy}
\end{figure*}

\begin{figure*}
\includegraphics[width=\textwidth]{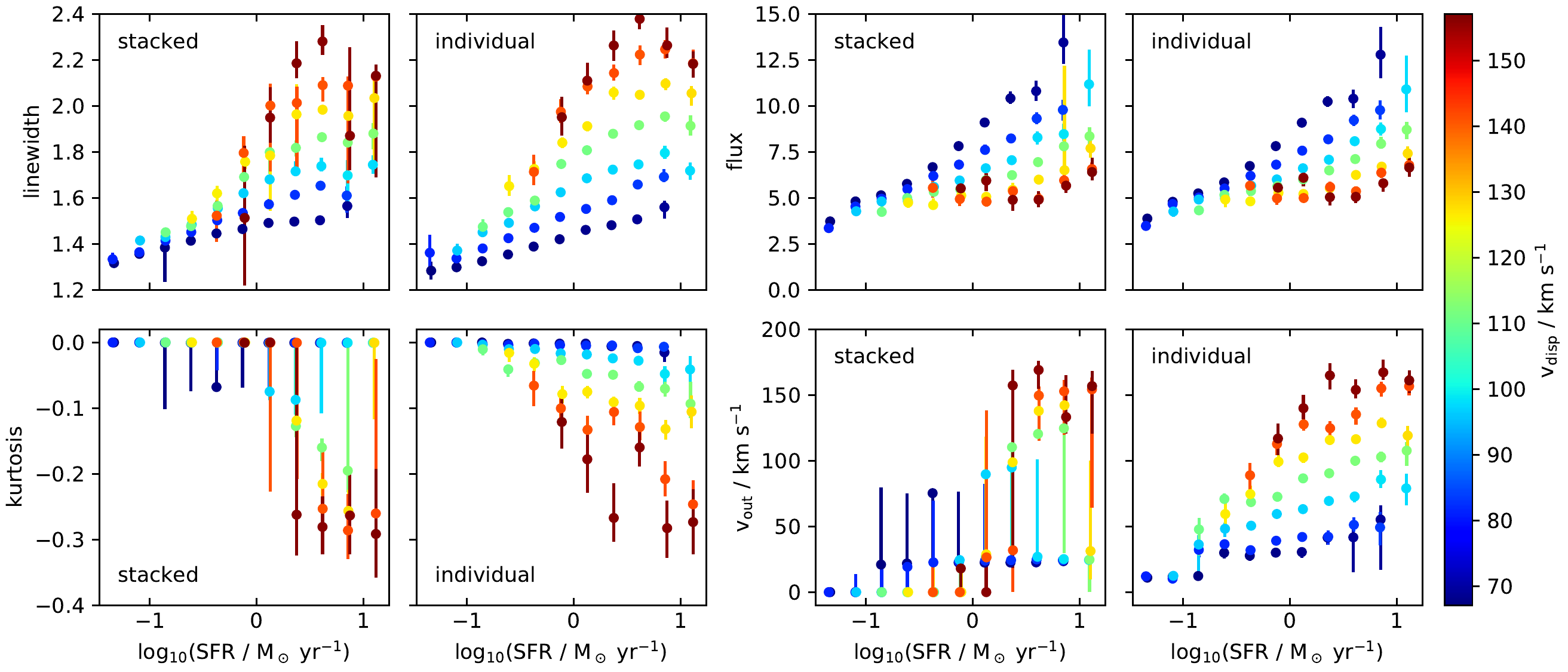}
\caption{Line properties of H$\beta$ from stacked spectra (equivalent to those in Figure~\ref{fig:sfr}) and individual galaxies across the parameter space in Figure~\ref{fig:param_sfr}, where the scatter points are colour-coded based on the $v_{\rm disp}$ value. The top-left, top-right, bottom-left, and bottom-right panels show the linewidth $\sigma$, line amplitude $A$, kurtosis $\kappa$, and outflow velocity $v_{\rm out}$, respectively. Both results are in general consistent with each other, and their error bars are estimated using a bootstrapping technique. The trends from individual galaxies are smoother with smaller error bars, but the line model is less sophisticated due to a limited number of spectral points.}
\label{fig:individual}
\end{figure*}

The kurtosis of emission lines was also investigated by \cite{Chen2016}, who focused on the H$\alpha$ emission line in disc star-forming galaxies. They found that the fraction of individual galaxies which contain platykurtic H$\alpha$ line profiles increased monotonically with increasing $M_\star$ primarily and with increasing SFR secondarily. This is consistent with our results in Section~\ref{sec:SFR} and Figure~\ref{fig:sfr}. However, they also found that such fraction was much larger in edge-on galaxies compared to face-on galaxies, which disagrees with our results in Section~\ref{sec:expAB} and Figure~\ref{fig:expAB}. To investigate this further, we repeated our analysis on each individual galaxy. The result is shown in Figure~\ref{fig:boxy}, where we divide the whole sample of individual galaxies into bins of $b/a$ and $v_{\rm disp}$ simultaneously. The left panel shows the population distribution with respect to both parameters, and the right panel shows the fraction, $F$, of galaxies with H$\beta$ kurtosis $\kappa<-0.5$. This reveals that the platykurtic H$\beta$ line profile is significantly more common in face-on galaxies (large $b/a$ value). However, the white numbers at the bottom of the right panel shows the overall $F$ across the columns, where the smallest value is given by the column with the largest $b/a$. This is because the galaxy population is dominated by galaxies with low $v_{\rm disp}$, which misrepresent the overall $F$. This demonstrates the necessity to take into account that each $b/a$ column is combined with galaxies of various $v_{\rm disp}$, which must be disentangled to properly examine how these galactic properties separately affect the properties of outflows.

\subsection{Analysing individual galaxies}
\label{sec:individual}

The analysis for H$\beta$ emission line in individual galaxies which has been used to generate Figure~\ref{fig:boxy} can be applied to complement our results for stacked spectra in Section~\ref{sec:result}. Here, we instead divide the whole sample of individual galaxies into bins of $v_{\rm disp}$ and SFR according to Figure~\ref{fig:param_sfr}. The H$\beta$ results from stacked galaxies and individual galaxies are compared and shown in Figure~\ref{fig:individual}, where the linewidth ($\sigma$), flux ($A$), kurtosis ($\kappa$), and outflow velocity ($v_{\rm out}$) are plotted in the top-left, top-right, bottom-left, and bottom-right panels, respectively. Overall, both results are consistent with each other as they should be, except for the minor differences in $\kappa$ and therefore $v_{\rm out}$. As demonstrated by Figure~\ref{fig:boxy} and the results from \cite{Chen2016}, the line profile of most galaxies is Gaussian-like ($\kappa=0$), and a platykurtic line profile can only be found in a smaller fraction of galaxies. For this reason, while the value of $\kappa$ across the individual galaxies averages to some non-zero value, the line profile obtained from stacked spectra may still yield $\kappa=0$. In general, the trends from individual galaxies are smoother with smaller error bars. However, since interpolation is not a viable option for spectra of individual galaxies, the complexity of the line model is limited by the number of spectral points. In this case, the number of spectral points was not enough to introduce an additional line component to fit the skewness, so we could only fit the H$\beta$ line with no more than a convolved line profile $F(\lambda)$.

\section{CONCLUSIONS}
\label{sec:conclude}

We analyse in this paper the profiles of prominent emission lines in a sample of $\sim53000$ star-forming galaxies ($\texttt{bptclass}=1$) from SDSS DR14. The galaxies are binned into 60 groups according to their stellar velocity dispersion ($v_{\rm disp}$) and star formation rate (SFR) to generate stacked spectra of their emission lines. Our analysis shows that most of the emission lines (H$\beta$, H$\alpha$, [\textsc{N\,ii}], [\textsc{S\,ii}]) are systematically platykurtic ($\kappa<0$) in galaxies with a higher mass and SFR. The radial outward motion in an outflow would cause the emission lines from the gas to appear platykurtic. Thus, these emission lines are expected to originate from the outflowing gas. In contrast, the [\textsc{O\,iii}] $\lambda$5007 line remains mesokurtic ($\kappa=0$). This can be explained if the emission region of the [\textsc{O\,iii}] line is not in the outflow, as expected from the different physical properties needed in gas with prominent [\textsc{O\,iii}] emission. When each group of galaxies is split into two subgroups according to their axial ratio, 4000\AA\  break and specific SFR, we find that the outflow signature is significantly stronger in face-on galaxies and in older galaxies. As line profiles trace gas kinematics along the line of sight, this result is consistent with the picture of star-formation activity driving galactic outflows along the rotation axis, which can be more easily detected in face-on galaxies. Galactic outflows exert an accumulated effect on the gas component, and hence older galaxies, which have sustained a longer total duration of star formation, show more noticeable outflow signatures.

\begin{acknowledgement}
This work is supported in part by a STFC Consolidated Grant awarded to UCL-MSSL. JA is supported by a STFC studentship. J. A. acknowledges financial support from INAF-WEAVE funds, program 1.05.03.04.05 and INAF-OABrera funds, program 1.05.01.01. IF acknowledges support from the Spanish Research Agency of the Ministry of Science and Innovation (AEI-MICINN) under the grant with reference PID2019-104788GB-I00. Funding for SDSS-III has been provided by the Alfred P. Sloan Foundation, the Participating Institutions, the National Science Foundation, and the U.S. Department of Energy Office of Science. The SDSS-III web site is http://www.sdss3.org/. This research has made use of NASA’s Astrophysics Data Systems.
\end{acknowledgement}

% PASA uses footnotes, not endnotes. \endnote in this template will behave like \footnote; and \printendnotes will not output anything.
% \printendnotes

\bibliography{lineshape}

\appendix

\section{Effects of potentially inaccurate stellar continuum subtraction on the line kurtosis}\label{sec:app1}

\begin{figure*}
\includegraphics[width=\textwidth]{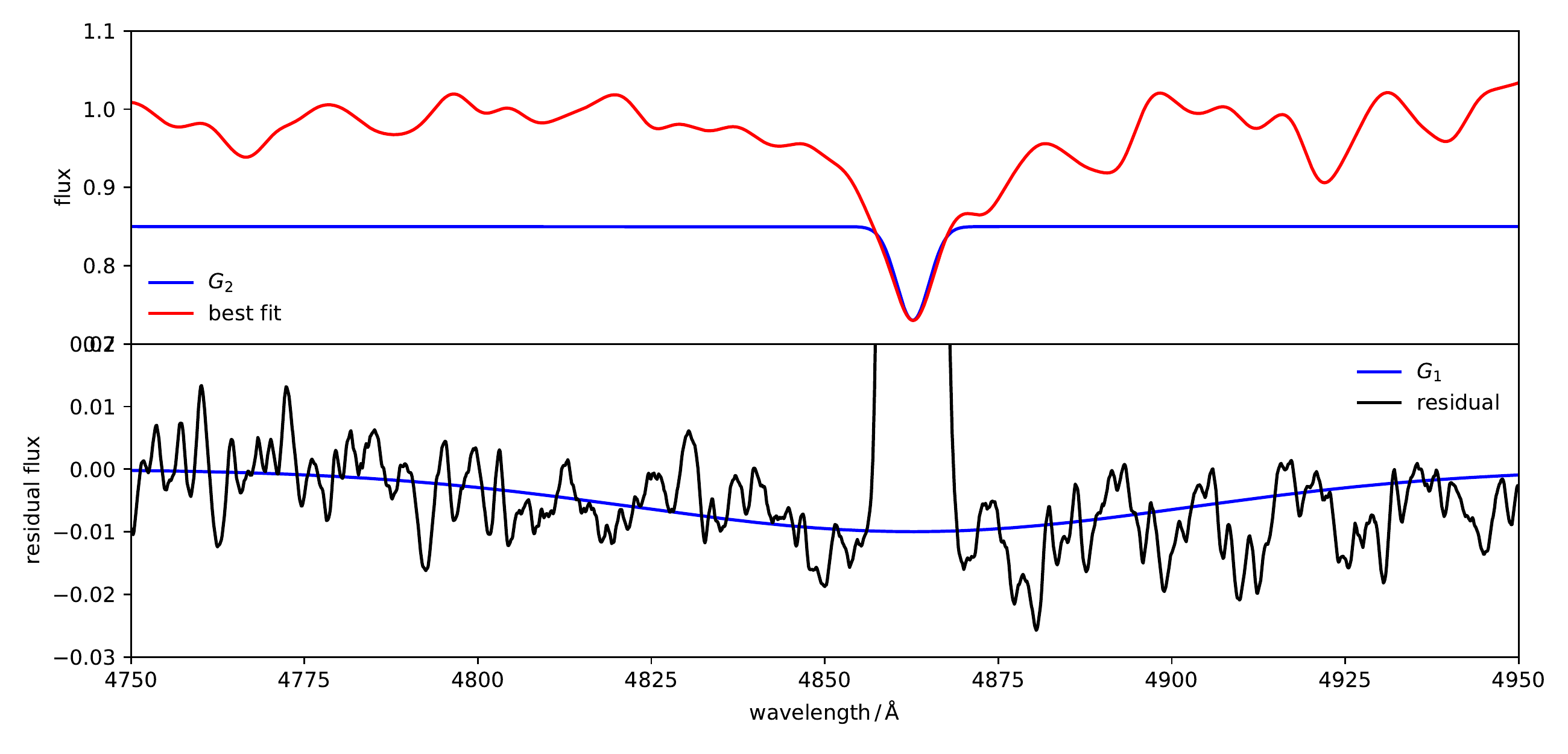}
\caption{Zoomed-in version of Figure~\ref{fig:ppxf} over the wavelength range $4750\,\textup{\AA}\le\lambda\le4950\,\textup{\AA}$. The best-fit of the stellar continuum using the pPXF algorithm is plotted in red in the upper panel, and the residual flux is plotted in black in the bottom panel. $G_1$ and $G_2$ are plotted in blue in the lower panel and upper panel respectively to illustrate the possible sources of error in fitting the stellar continuum.}
\label{fig:1}
\end{figure*}

Here, we test whether the trend found regarding a platykurtic H$\beta$ line profile may suffer from a systematic bias due to an inaccurate stellar continuum subtraction in the H$\beta$ wavelength region. To investigate this problem, we created a zoomed-in version of Figure~\ref{fig:ppxf}. As shown in the bottom panel of Figure~\ref{fig:1}, the residual flux near the wings of the H$\beta$ line lies more often below zero. This may be interpreted as an over-subtraction which spans roughly 100 \AA{}. Does this potential issue affect our results? To account for this, we assigned a Gaussian $G_1(\lambda)$ with amplitude $A=1$, mean $\lambda_0=4862.68$ \AA{}, and standard deviation $\sigma=40$ \AA{} to compensate for any potential difference between the observed flux and the best-fit. We repeat the line-fitting procedure for $F(\lambda)$, $F(\lambda)+G_1(\lambda)$, and $F(\lambda)-G_1(\lambda)$, and the results are
\begin{align*}
\begin{split}
A=5.585,\ \lambda_0=4862.59\,\textup{\AA},\ \sigma=1.35\,\textup{\AA}, 
 v_{\rm out}=171.17\,
 {\rm km\;\!s}^{-1}
 %\rm km/s
 \ ,  \\
{\rm for}\ F(\lambda),
\end{split}\\
\begin{split}
A=5.586,\ \lambda_0=4862.59\,\textup{\AA},\ \sigma=1.35\,\textup{\AA}, 
 v_{\rm out}=171.17\,
 {\rm km\;\!s}^{-1}
 %\rm km/s
 \ ,  \\
{\rm for}\ F(\lambda)+G_1(\lambda),
\end{split}\\
\begin{split}
A=5.584,\ \lambda_0=4862.59\,\textup{\AA},\ \sigma=1.35\,\textup{\AA}, 
 v_{\rm out}=171.17\,
 {\rm km\;\!s}^{-1}
 %\rm km/s
 \ , \\
{\rm for}\ F(\lambda)-G_1(\lambda),
\end{split}
\end{align*}
which shows that the results regarding the shape of the emission line, and the corresponding $v_{\rm out}$ are unaffected. This is within our expectations, because kurtosis is only concerned with the shape of the narrow emission line. Any large-scale spectral fluctuation in the shape of the stellar continuum is rendered flat in the proximity of the H$\beta$ emission line that spans only a few \AA{}. Note that a model that assumes an extended pair of high-velocity tails may include two Gaussian components with the same mean, and the second one having larger width corresponding to the outflow. Such a model would give positive kurtosis, accounting for the extended wings. However, this may be overly simplistic if the two components have different means, i.e. different line-of-sight velocities relative to the rest frame of the galaxy, which can be invoked to produce negative kurtosis \citep[see, e.g.][]{Westmoquette2011}.

The width of the H$\beta$ absorption line is characterised by the stellar velocity dispersion, $v_{\rm disp}$. Since the galaxies for this stacked spectrum are chosen within the $135\,{\rm km/s}\le v_{\rm disp}\le150\,{\rm km/s}$ interval, we can take the average value that translates into $\sigma=2.31\,\textup{\AA}$. We assign another Gaussian $G_2(\lambda)$ with $A=0.7$, $\lambda_0=4862.68$ \AA{}, and $\sigma=2.31$ \AA{} to account for any difference between the real and fitted H$\beta$ absorption line, as shown in the top panel of Figure~\ref{fig:1}. We repeat the line-fitting procedure for $F(\lambda)$, $F(\lambda)+G_2(\lambda)$, and $F(\lambda)-G_2(\lambda)$, and the results are
\begin{align*}
\begin{split}
A=5.585,\ \lambda_0=4862.59\,\textup{\AA},\ \sigma=1.35\,\textup{\AA}, v_{\rm out}=171.17\,
{\rm km\;\!s}^{-1}
%\rm km/s
\ ,\\
{\rm for}\ F(\lambda),
\end{split}\\
\begin{split}
A=6.269,\ \lambda_0=4862.60\,\textup{\AA},\ \sigma=1.40\,\textup{\AA}, v_{\rm out}=169.27\,
{\rm km\;\!s}^{-1}
%\rm km/s
\ ,\\
{\rm for}\ F(\lambda)+G_2(\lambda),
\end{split}\\
\begin{split}
A=4.908,\ \lambda_0=4862.58\,\textup{\AA},\ \sigma=1.31\,\textup{\AA}, v_{\rm out}=174.49\,
{\rm km\;\!s}^{-1}
%\rm km/s
\ ,\\
{\rm for}\ F(\lambda)-G_2(\lambda),
\end{split}
\end{align*}
where the effect on the shape of the emission line is negligible, because the strength of the narrow H$\beta$ emission line is overwhelmingly greater than that of the H$\beta$ absorption line in a starburst system.

%Note that these scenarios ($G_1$ and $G_2$) are greatly exaggerated for demonstration purposes.
Note that the amplitudes of $G_1$ and $G_2$ are significantly greater than the 1$\sigma$ error of these amplitudes.
The actual fluctuation or inaccuracy is expected to be substantially smaller. The robustness of the continuum fit is also tested by using another library of SSP spectra, the E-MILES model, and the results are completely consistent, proving that the process of removing the stellar continuum is robust.

\end{document}